\documentclass[conference,compsoc]{IEEEtran}
\usepackage{graphicx} 
\usepackage{xspace}             
\usepackage{microtype}
\usepackage{subfigure}
\usepackage{booktabs} 
\usepackage[OT1]{fontenc}
\usepackage{multirow}
\usepackage{adjustbox}
\usepackage{tabularray}
\usepackage{algpseudocode}
\usepackage{comment}
\usepackage{amsthm}
\newtheorem{prop}{Proposition}
\usepackage{algorithm}
\usepackage{algpseudocode}
\usepackage{amsmath}
\usepackage{amsfonts}
\usepackage{lineno}
\PassOptionsToPackage{table}{xcolor}
\usepackage[table,xcdraw]{xcolor}
\definecolor{mygray}{gray}{0.85}

\usepackage[utf8]{inputenc}
\usepackage[OT1]{fontenc}
\usepackage[russian, english]{babel}
\usepackage{pgfplots}

\usepackage{mathtools}
\usepackage[]{mdframed}
\usepackage{subcaption}
\usepackage{threeparttable}

\usepackage{tikz}
\usetikzlibrary{tikzmark,fit,calc,positioning,arrows,graphs,shapes.misc}
\usepackage{fontawesome}
\usepackage[normalem]{ulem}
\usepackage[hyphens]{url}
\usepackage{hyperref}
\usepackage{threeparttable}
\usepackage{tablefootnote}
\usepackage[symbol*]{footmisc}

\hypersetup{
    colorlinks=true,
    linkcolor=violet,
    urlcolor=cyan,
    citecolor=black,
    pdfpagemode=FullScreen,
}
\usepackage[capitalize,nameinlink]{cleveref}

\algnewcommand{\LineComment}[1]{\State \(\triangleright\) #1}

\newcommand{\argmax}[1]{{\underset{#1}{\mathrm{argmax}}}}
\newcommand{\argmin}[1]{{\underset{#1}{\mathrm{argmin}}}}

\newcommand{\ie}{\textit{i.e.,}\@\xspace}
\newcommand{\eg}{\textit{e.g.,}\@\xspace}
\newcommand{\etal}{\textit{et al.}\@\xspace}

\renewcommand{\eqref}[1]{\mbox{Eq.~\ref{#1}}}

\usepackage{siunitx}
\usepackage{acronym}

\newcommand{\newparagraph}[1]{\smallskip\noindent \textbf{#1}}

\makeatletter
\pgfplotsset{
    /tikz/max node/.style={
        anchor=south
    },
    /tikz/min node/.style={
        anchor=north
    },
    mark min/.style={
        point meta rel=per plot,
        visualization depends on={x \as \xvalue},
        scatter/@pre marker code/.code={%
            \ifx\pgfplotspointmeta\pgfplots@metamin
                \def\markopts{}%
                \node [min node] {
                    \pgfmathprintnumber[fixed]{\xvalue},%
                    \pgfmathprintnumber[fixed]{\pgfplotspointmeta}
                };
            \else
                \def\markopts{mark=none}
            \fi
            \expandafter\scope\expandafter[\markopts,every node near coord/.style=green]
        },%
        scatter/@post marker code/.code={%
            \endscope
        },
        scatter,
    },
    mark max/.style={
        point meta rel=per plot,
        visualization depends on={x \as \xvalue},
        scatter/@pre marker code/.code={%
        \ifx\pgfplotspointmeta\pgfplots@metamax

                \def\markopts{}%
                \node [max node] {
                    \tiny\pgfmathprintnumber[verbatim]{\pgfplotspointmeta}
                };

        \else
        \fi
            \expandafter\scope\expandafter[\markopts]
        },%
        scatter/@post marker code/.code={%
            \endscope
        },
        scatter
    }
}
\makeatother

\newlength{\saveparindent}
\newlength{\saveparskip}
\newenvironment{newitemize}{%
\begin{list}{\mbox{}\hspace{5pt}$\bullet$\hfill}{\labelwidth=15pt%
\labelsep=4pt \leftmargin=12pt \topsep=0pt%
\setlength{\listparindent}{\saveparindent}%
\setlength{\parsep}{\saveparskip}%
\setlength{\itemsep}{1pt} }}{\end{list}}

\usepackage{listings}
\usepackage{color}

\definecolor{dkgreen}{rgb}{0,0.6,0}
\definecolor{gray}{rgb}{0.5,0.5,0.5}
\definecolor{mauve}{rgb}{0.58,0,0.82}
\lstdefinestyle{myListingStyle} 
    {
        basicstyle = \small\ttfamily,
        breaklines = true,
    }

\usepackage{xcolor}
\definecolor{blue}{RGB}{0,0,255}
\definecolor{teal}{RGB}{0,128,128}
\definecolor{purple}{RGB}{128,0,128}
\definecolor{green}{RGB}{0,128,0}
\definecolor{violet}{RGB}{148,0,211}

\usepackage[encapsulated]{CJK}

\begin{document}

\title{\Large \bf Fun-tuning: Characterizing the Vulnerability of Proprietary LLMs to Optimization-based Prompt Injection Attacks via the Fine-Tuning Interface}

\author{
\IEEEauthorblockN{Andrey Labunets\textsuperscript{1}, Nishit V. Pandya\textsuperscript{1}, Ashish Hooda\textsuperscript{2}, Xiaohan Fu\textsuperscript{1}, Earlence Fernandes\textsuperscript{1}}
\IEEEauthorblockA{\textsuperscript{1}UC San Diego,
\textit{\{alabunets, nipandya, x5fu, efernandes\}@ucsd.edu}}
\IEEEauthorblockA{\textsuperscript{2}University of Wisconsin Madison,
\textit{ahooda@wisc.edu}}
}

\date{}
\maketitle

\begin{abstract}
    We surface a new threat to closed-weight Large Language Models (LLMs) that enables an attacker to compute optimization-based prompt injections. Specifically, we characterize how an attacker can leverage the loss-like information returned from the remote fine-tuning interface to guide the search for adversarial prompts. The fine-tuning interface is hosted by an LLM vendor and allows developers to fine-tune LLMs for their tasks, thus providing utility, but also exposes enough information for an attacker to compute adversarial prompts. Through an experimental analysis, we characterize the loss-like values returned by the Gemini fine-tuning API and demonstrate that they provide a useful signal for discrete optimization of adversarial prompts using a greedy search algorithm. Using the PurpleLlama prompt injection benchmark, we demonstrate attack success rates between 65\% and 82\% on Google's Gemini family of LLMs. These attacks exploit the classic utility-security tradeoff --- the fine-tuning interface provides a useful feature for developers but also exposes the LLMs to powerful attacks.
\end{abstract}

\section{Introduction}

Large Language Models (LLMs) face numerous security and privacy issues, such as being forced to output text that violates a vendor's policies~\cite{rao2023tricking,liu2023jailbreaking,jain2023baseline,wei2023jailbroken,yu2023gptfuzzer, niu2024jailbreaking, jiang2024artprompt} or being tricked into misusing its access to tools via prompt injection attacks~\cite{greshake2023youve,liu2023prompt,markdownattack,zhan2024injecagent, yi2023benchmarking,emailAgentHijacking}. The community's ultimate goal is to create secure and private LLMs and an important step along the way is to thoroughly explore the novel attack vectors that the models might face. We contribute to this line of work and surface a new attack vector on LLMs that allows an attacker to compute optimization-based prompt injections for closed-weights proprietary LLMs.

Specifically, we demonstrate how an attacker can (mis)use the fine-tuning interface to an LLM to guide the search for adversarial prompts. Many vendors allow consumers to remotely fine-tune closed-weights models to specialize them for various downstream tasks~\cite{googleFinetuning,gptfinetuning,aws2023fine}. The vendor itself supports the fine-tuning task using an extensive infrastructure involving datacenters of GPUs. Fine-tuning interfaces return training progress metrics across a user-supplied training and validation dataset. Our core insight is that by setting a very small learning rate, an attacker can obtain a signal that approximates the log probabilities of target tokens (``logprobs'')  for the LLM. As we experimentally show, this allows attackers to compute graybox optimization-based attacks on closed-weights models. Using this approach, we demonstrate, to the best of our knowledge, the first optimization-based prompt injection attacks on Google's Gemini family of LLMs (see \cref{fig:example-attack-password}).

Our attacks exploit a fundamental trade-off between security and utility. Vendors want to expose a fine-tuning interface so that developers can gain the benefits of specialized models, and thus, they must expose fine-grained training metrics so that the developers can do an effective job at fine-tuning the models. Our work shows that this utility is fundamentally at odds with security --- the loss-like training metrics that are useful for benign fine-tuning usage are also helpful to attackers who can guide their search for adversarial prompts.

\begin{figure}[t]
    \centering
    \begin{adjustbox}{width=\linewidth}
            \input{plots/passwod_attack_2}
    \end{adjustbox}
    \caption{Example prompt injection with our method on Gemini 1.5 Flash (taken from PurpleLlama benchmark). Our attack uses fine-tuning loss data to compute a payload (shown in red) that wraps an existing prompt injection trigger (bolded) to ``boost'' it. This forces the model to obey the injected instructions. The payload and the instructions remain as a single-line comment, preserving Python syntax.}
    \label{fig:example-attack-password}
\end{figure}

Existing adversarial prompting techniques fall into two categories: (1) Linguistic attacks that rely on rephrasing text prompts until they achieve a malicious goal, mostly to ``jailbreak'' the model by getting it to respond with text that violates a content policy~\cite{andriushchenko2024jailbreakingleadingsafetyalignedllms,githubChatGPTDanJailbreakmd,chao2024jailbreakingblackboxlarge,mehrotra2023treeOfAttacks}; (2) Automated optimization-based search techniques that rely on loss data to guide a discrete search for tokens that achieve the attacker's goals~\cite{zou2023universal}. Our work lies in the second category of attacks. The key differentiating factor between automated attack types is the level of access an attacker has to loss information. For example, whitebox attacks utilize loss combined with gradients~\cite{zou2023universal,pasquini2024neural}, graybox attacks utilize logprobs from the inference endpoints~\cite{hayase2024query,sitawarin2024palproxyguidedblackboxattack}; and blackbox attacks rely only on the textual output from the model~\cite{chao2024jailbreakingblackboxlarge,liu2024autodangeneratingstealthyjailbreak}. LLM vendors have recently mitigated some of the graybox attack vectors by either not returning logprobs as part of a inference call, or returning only a small subset of logprobs~\cite{openaiapireference,googleGeneratingContent,carlini2024stealingproductionlanguagemodel}. In both cases, there is not sufficient information to effectively guide the search for an adversarial prompt. Our work shows that existing greedy search algorithms can be adapted to use loss-like information from the fine-tuning interface to create attacks.

In our \textit{fun-tuning} attacks, the attacker runs a single iteration of fine-tuning with a very small learning rate. The small learning rate ensures that the base model does not change significantly, subverting the point of fine-tuning, which is to update the model's weights. The output of this operation is a loss-like value representing how far the model's true output is from the desired output. We use this loss-like value to guide our search procedure. There are two technical challenges in using the fine-tuning functionality this way.

The first challenge is establishing that the fine-tuning loss is a good signal to guide the search for adversarial prompts. The fine-tuning interfaces of commercial vendors are closed source and the documentation does not provide details. Using a range of experiments, we partially reverse-engineer the loss function used by Google's Gemini and empirically establish that the reported training loss is a noisy signal that can be useful for discrete optimization of prompts.

The second challenge is that fine-tuning interfaces randomize the order of items in the training and validation sets. The losses that get reported are a permutation of the input training data order. This is problematic because the attacker has to systematically evaluate the effect of different token substitutions when creating the adversarial prompt. We address this challenge using an approximate de-randomization procedure that uses a carefully crafted training dataset to approximate the permuted training order. We validate the utility of our approximate procedure by comparing it with another inefficient, but provably-correct, method and find that our procedure approximates the true permutation within reasonable bounds.

We focus on creating prompt injection attacks because they are the most realistic security problem faced by systems built using large models. These systems, or agents, can use tools to solve a variety of tasks such as automatically finding bugs in software and managing calendars and emails. A prompt injection attack in the context of an agent-based system can affect the confidentiality and integrity of user data. Consider a simple agent that has access to a user's email. An attacker could craft an adversarial prompt and send an unsolicited email to a victim user. If the user asks their email-handling agent to ``summarize my latest email,'' the adversarial prompt will hijack the agent and cause it to also send out the user's emails to the attacker, thus violating confidentiality~\cite{emailAgentHijacking,greshake2023youve}.

Mitigating the fun-tuning attack vector is non-trivial because it exploits the utility-security trade-off. Fine-grained control over training hyperparameters (learning rate, batch size, epochs, randomization seed for training set shuffling) is crucial to developers who are using the interface in benign ways, but as we show, it is also helpful to attackers. Therefore, any changes that reduce control over hyperparameters (e.g., setting a large minimum value on learning rate) can negatively affect benign developers and the utility of the fine-tuning interface itself. Scanning the training set for the presence of malicious data before running the fine-tuning job is a potential mitigation (\eg to look for content policy violations), but can be evaded using encoding techniques that conceal the semantics of the data~\cite{halawi2024covert} and is not always possible for prompt injection attacks. 

\newparagraph{Contributions.}
\begin{newitemize}
    \item \textbf{New Attack Surface Characterization.} We surface and experimentally characterize a new attack vector on LLM systems that exploits a fundamental tension between utility and security. Specifically, we show how an attacker can misuse the fine-tuning interface to create adversarial prompts. We call these \textit{fun-tuning} attacks. We solve two technical challenges along the way: (a) establishing that the fine-tuning loss is suitable for discrete optimization of prompt injections and (b) derandomizing the reported loss-like values to obtain a usable signal for adversarial prompt optimization.

    \item \textbf{Experimental Analysis of Optimization-based Prompt Injections.} We experimentally evaluate the vulnerability of Google's Gemini model series to fun-tuning attacks. We show that existing discrete optimization algorithms for prompt injections can be modified to incorporate fine-tuning losses for guidance. Using the popular prompt injection benchmark PurpleLlama~\cite{bhatt2023purplellamacybersecevalsecure}, we show attack success rates between 65\% and 82\% for Google's Gemini family of models. We also find that the attacks transfer between various Gemini models with relatively high success rates. 
    
\end{newitemize}

\noindent
\newparagraph{Disclosure and Ethics.} We disclosed the issue to Google on November 18, 2024. Google deployed the following mitigation in early April 2025: ``We constrained the API parameters that they were relying on. In particular, capping the learning rate to a value that would rule out small perturbations and limiting the batch size to a minimum of 4, such that they can no longer correlate the reported loss values to the individual inputs.'' Our goal with this work is to raise awareness and begin a conversation around the security of fine-tuning interfaces and their role in helping create prompt injection attacks on closed-weight models. We conducted all experiments using the standard developer fine-tuning interface. Although we created attacks, the fine-tuning job and its impact on the provider are indistinguishable from benign fine-tuning jobs. We did not deploy any of these attacks in the wild. Code is available at \url{https://github.com/earlence-security/fun-tuning}.
\section{Background}
We model an LLM $\mathcal{M}_{\Theta}$ as a probability distribution over the next token conditioned on its input tokens, that is, for an input sequence of tokens $x_{1:n} = (x_1, x_2, \dots, x_n)$, $\mathcal{M}_{\Theta}$ outputs a probability distribution $P(y|x_{1:n}; \Theta)$, where $\Theta$ denotes the model parameters, $y$ represents the next token to be generated. Here, all tokens come from a discrete set $V = \{T_1, T_2, ..., T_{|V|}\}$ (called the vocabulary of the LLM).

\newparagraph{Using LLMs in Generative/Inference Mode.} The next output token $y_1$ is generated by sampling from the probability distribution $y_1 \sim P(y|x_{1:n}; \Theta)$ according to some sampling procedure.
In practice, LLMs generate a sequence of tokens until a special ``end-of-text" token is generated. Therefore, on input $x_{1:n} \in V^*$, an LLM will return an output sequence $y_{1:m} \in V^*$ with probability 
\begin{equation}
    \label{eq:1}
    P(y_{1:m}\vert x_{1:n}; \Theta) = \Pi_{i=1}^{m}P(y_i \vert x_{1:n}, y_{1:i-1}; \Theta)
\end{equation}

In addition to the generated stream of tokens, LLM APIs return a variety of additional information related to the probability of generation of the output in the form of "logprobs".
Logprobs of a token $y$ represent the probability of generation of that token $y$ given the input $x = x_{1:n}$.
That is for a sequence of tokens $y = y_{1:m}$, we have
\begin{equation}      \label{eq:logprobs_definition}
    \mathrm{Logprobs}(y\vert x; \Theta)
    = -\sum\limits_{i=1}^{m} \mathrm{log}P(y_i \vert x, y_{1:i-1}; \Theta)
\end{equation}

Some APIs (\eg Cohere, GooseAI) return the complete logprobs vector, \ie the logprobs for each token in the vocabulary. Some APIs return only the $\mathrm{Top}$-N logprobs (e.g., OpenAI returns top-20), and others do not return any logprobs information at all. A longer list can be found in Table 6 of~\cite{sitawarin2024pal}.

\newparagraph{Training.}
Generative LLMs are trained by finding parameters $\Theta$ which minimize the Cross-entropy loss between the predicted probability distribution over the next token $\mathcal{M}_{\Theta}(x^{i}) = P(y | x^{i})$ and the true next token $y^{i}$ summed over the training dataset $I$, where $x^{i}$ denotes $i$'th training example of $I$.
\begin{equation}
    \label{eq:2}
    \hat{\Theta} = \argmin{\Theta} \sum_{i \in I} \mathrm{CrossEntropy}(\mathcal{M}_{\Theta}(x^{i}), y^{i})
\end{equation}

Expanding the definition of Cross-entropy loss, we get
\begin{equation}
    \label{eq:3}
    \hat{\Theta} = \argmin{\Theta} \big( - \sum_{i \in I} \log P(y^{i} | x^i; \Theta) \big)
\end{equation}

\newparagraph{Prompt Injections.}
In real world systems, the input to an LLM usually consists of system and user prompts, separated by special tokens and concatenated into a single token sequence, often called a conversation.
For simplicity, we assume an LLM prompt is a pair consisting of a system prompt $x_{\mathrm{Sys}}$ and a user prompt $x_{\mathrm{User}}$. Crucially, the system prompt and user prompt come from sources with different security contexts and assume a trust boundary between them: one of these values might be controlled adversarially.

Prompt Injection attacks assume partial control of an LLM input and use it to achieve objectives not originally intended by the developer or the actual user.
Prompt Injections can lead to several real-world consequences such as leakage of private data and tool misuse~\cite{greshake2023youve,liu2023prompt,markdownattack,zhan2024injecagent,emailAgentHijacking}. This compromises the confidentiality and integrity of user resources connected to the LLM.

A prompt injection attack occurs when a trusted input (such as the system prompt of an LLM agent and the user prompt), concatenated with an untrusted malicious input, (such as text retrieved from a third-party webpage to be summarized), causes the LLM to deviate from its original task and follow the instructions of the malicious input. Formally, we define prompt injection when the following holds:
\begin{equation}
    \label{eq:10}
    LLM(x_{\mathrm{Trusted}} \Vert x_{\mathrm{Adv}}) \approx LLM(x_{\mathrm{Adv}}) \approx y_{Target}
\end{equation}

The task of an attacker is to craft an adversarial input $x_{\mathrm{Adv}}$ such that, if combined with $x_\mathrm{Trusted}$, the LLM outputs a sequence of tokens $y_{\mathrm{Target}}$ of the attacker's choice instead of the intended output $y_{\mathrm{True}}$.
The definition might be extended to assume $LLM(x_{\mathrm{Adv}})$ is still conditioned on $x_{\mathrm{Trusted}}$ (but no longer instructed by it) to cover the cases where $x_{\mathrm{Adv}}$ instructs $LLM$ to reveal $x_{\mathrm{Trusted}}$, but we leave it out of scope for simplicity.

We also note that, in general, the adversarial input could be combined with the trusted input in several different ways - such as concatenation of strings, parameterized queries wrapping each prompt with special characters, interspersed with each other or via further pre-processing \cite{chen2024struqdefendingpromptinjection}. However, for simplicity and without loss of generality, we assume that the trusted and untrusted sequences are combined using plain concatenation denoted by $\Vert$.
Under this setting, the adversary needs to find $x_{\mathrm{Adv}}$ such that
$$LLM(x_{\mathrm{Trusted}} \Vert x_{\mathrm{Adv}}) = y_{\mathrm{Target}}$$

Most early prompt injection attacks were crafted manually and exploited various model-specific quirks~\cite{markdownattack,greshake2023youve,emailAgentHijacking}.
These early attacks benefit from linguistic approaches and automated rephrasing methods, but here we formalize the problem mathematically.

The problem of finding such an $x_{\mathrm{Adv}}$ can be formulated as an optimization problem where the objective is to find an $x_{\mathrm{Adv}}$ which maximizes the probability of the LLM outputting the string $y_{\mathrm{Target}}$, \ie
\begin{equation}
    \label{eq:4}
    x_{\mathrm{Adv}} = \argmax{x}~P(y_{\mathrm{Target}} \vert (x_{\mathrm{Trusted}} \Vert x); \Theta)
\end{equation}

This is equivalent to finding an input $x_{\mathrm{Adv}}$ that minimizes the cross-entropy loss over the target string: 
$$
    x_{\mathrm{Adv}} = \argmin{x}\ \mathrm{CrossEntropy}(\mathcal{M}_{\Theta}(x_{\mathrm{Trusted}} \Vert x), y_{\mathrm{Target}})
$$
or equivalently
\begin{equation}
    \label{eq:min_cross_entropy_loss}
    x_{\mathrm{Adv}} = \argmin{x}\ -\mathrm{Logprobs}(y_{\mathrm{Target}} \vert x_{\mathrm{Trusted}} \Vert x; \Theta)
\end{equation}

The above discrete optimization problem was addressed in the white-box setting using the Greedy Coordinate Gradient algorithm which relies on being able to compute gradients that guide a search for an adversarial input \cite{zou2023universal}. In the graybox and blackbox setting, where gradients are not available, prior optimization based approaches rely either on transferability or on being able to compute the logprobs for a target token of the attacker's choice \cite{sitawarin2024palproxyguidedblackboxattack,hayase2024query}.

\newparagraph{LLM Fine-Tuning Functionality.}
Fine-tuning is a procedure that allows users to further train a pre-trained base model on custom or proprietary data.
Fine-tuning of an LLM can help with increasing model accuracy and reducing hallucinations on domain-specific knowledge, and allow for simplified prompts~\cite{google2022finetuningeffect,zhang2024scalingmeetsllmfinetuning}.
Several AI companies such as OpenAI, Google, and Amazon provide the fine-tuning functionality as a service for users~\cite{googleFinetuning,gptfinetuning,aws2023fine}.

To use the fine-tuning functionality, users need to prepare and format a high-quality dataset which is reflective of their use case and fine-tune a base model on that dataset with an appropriate fine-tuning configuration. The fine-tuned model is then made available for inference for the custom use-case.

For example, we provide a high level description of the LLM fine-tuning API provided by Google AI Studio (``Gemini API").
The Gemini Fine-Tuning API accepts a base model and a training dataset in the form of a list of input-output pairs of strings.
In addition, it lets users specify the number of epochs for the fine-tuning, the batch size, the learning rate.
The Fine-Tuning API responds with the training loss for every iteration (step) of the optimization process.
Additionally, the API endpoint shuffles the dataset examples before training on them, so that the resulting losses come in a pseudorandom order.

Fine-tuning interfaces from different companies provide a similar level of control over the training hyperparameters (see \cref{sec:disc} for more details).
\section{Threat Model and Attack Constraints}
\label{sec:threat}
The attacker's goal is to create prompt injection attacks on a target LLM. They are a third party who wish to take control of an existing conversation, and then force the LLM to follow a different set of instructions. The attacker can deliver the prompt injection using a variety of ways, such as poisoning a webpage that the user might want summarized, sending an unsolicited email to the user's LLM-based agent, or modifying a code repository that the user might be analyzing~\cite{greshake2023youve,emailAgentHijacking,bhatt2023purplellamacybersecevalsecure}.

Successful prompt injection attacks must obey two properties. First, the domain of the LLM-based agent (e.g., code editing agent, email/calendar handler agent, web browsing agent) imposes constraints on the syntax and size (in number of tokens) of the attack prompt. For example, for an LLM-based coding assistant, the attack must exist as a valid comment and cannot break the syntax of the programming language. Similarly, if the user has delegated calendar management to the LLM, then the attack prompt must be delivered as a valid calendar event, limiting its size. Thus, for the generality of the prompt injection attacks that we create in this paper, we impose a size limit on the number of tokens that the attacker can inject.
Additionally, for code examples, we don't inject newline characters so that the adversarial prompt stays within a single commented out line.
The attack size varies based on the application domain --- 
we conduct our experiments with $97\%$ of all attack prompts sizes being below 100 tokens (all of them are shorter than 500 characters). Out of this, 40 tokens are the output of the optimization algorithm and the remaining tokens represent the attacker's instruction in natural language. The optimizer-controllable tokens are a configurable parameter of our algorithm. Thus, the final attack takes the form of a sequence of optimizer-created tokens sandwiching the attacker's natural language instruction (\cref{fig:example-attack-password}).

Second, the attack should be stealthy and must cause the target LLM to only produce the attacker's desired output without anything else. This is a stronger requirement compared to prior work in prompt injections~\cite{greshake2023youve, bhatt2023purplellamacybersecevalsecure}, but is one we believe to be an important constraint. Recent LLM products have been fine-tuned specifically to resist prompt injection attempts~\cite{wallace2024instructionhierarchytrainingllms}, which might result in a blocked request or might steer the model behavior towards returning a more truthful answer.
In that context, we observed that Gemini models often inform the user that its behavior is potentially overridden by a specific prompt when seeing some unusual requests. For example, Gemini in our experiments sometimes responded ``\textit{The code is designed to calculate the area of a circle with a radius of 5. However, the code has a comment that explicitly overrides the function's calculation and instructs the code to output `10'.}'' The attacker will not want the user to be aware that there is a prompt injection and thus we require our attacks to not create such warnings. 

We assume that the attacker has access to the fine-tuning interface for a target LLM. This type of access is graybox because the fine-tuning interface returns a loss signal. Most LLM vendors allow anyone to sign up and become a developer. 
\section{Experimental Analysis of the Gemini Fine-Tuning Interface}

The problem of generating adversarial inputs that force an LLM $\mathcal{M}_{\Theta}$ to output a string $y_{\mathrm{Target}}$ can be phrased as an optimization problem minimizing the unweighted cross-entropy loss over a given target string (\cref{eq:min_cross_entropy_loss}).
In absence of logprobs information, we can't directly compute the cross-entropy loss for a desired input and target string. To address this problem, we rely on the training losses reported by the fine-tuning API as a proxy loss function.

The key insight behind our approach is that for a small learning rate which is close enough to $0$, the parameters of a model should stay nearly constant. Therefore, if we send a fine-tuning request with a single training example with a single pair of input-output strings $(x_{\mathrm{Trusted}} \Vert x, y_{\mathrm{Target}})$ and fine-tune it for a single epoch with a very small learning rate, the loss metric reported should leak information about the cross-entropy loss of the target model. Thus, we can instead try to solve the following proxy optimization problem:
$$x_{\mathrm{Adv}} = \argmin{x}\ \mathrm{TrainingLoss}(\mathcal{M}_{\Theta}(x_{\mathrm{Trusted}} \Vert x), y_{\mathrm{Target}})$$

In this section, we focus on the Fine-Tuning API for their Gemini class of models and empirically establish that the training loss is a useful proxy for optimization. We do this in two steps:
\begin{newitemize}
    \item We probe the API to understand and validate our hypothesis that a small learning rate does not significantly affect the base model. This is important because even though the learning rate is an externally controllable hyperparameter, some fine-tuning interfaces treat these as \textit{multipliers} to an internal, hidden learning rate.
    \item We partially reverse-engineer the loss values reported by the API and establish that they are indeed a good proxy for discrete optimization.
\end{newitemize}

\subsection{Fine-tuning Hyperparameter Analysis}
\label{subsec:hyperparam_anal}

The Gemini Fine-tuning API accepts as input a training set, consisting of a list of pairs of input and output strings. Additionally, the API also allows users to control some hyperparameters for the fine-tuning process, specifically, the number of epochs, the batch size, and the learning rate. We probe the API and validate the following behavior of the Gemini Fine-tuning procedures.

\newparagraph{Small learning rates do not change training loss values significantly.} While the API only accepts values larger than roughly $10^{-45}$ as a valid learning rate, we find experimentally that for learning rate values between roughly $10^{-13}$ and $10^{-45}$, the loss values reported by the API (when controlled for the training set and other hyperparameters) stay constant (up to the precision of the numbers reported by the API) and are independent of the learning rate itself.
Concretely, we take a fixed training dataset $D$ consisting of $n$ distinct training examples and send it to the fine-tuning API with a learning rate $\alpha_0 \approx 10^{-45}$ and observe the losses returned, say $L^{(\alpha_0)} = \{l^{(\alpha_0)}_1, l^{(\alpha_0)}_2, \dots l^{(\alpha_0)}_n\}$. We then send $D$ again to the fine-tuning API, this time with a learning rate $\alpha_1 \approx 10^{-44}$ and collect the losses $L^{(\alpha_1)} = \{l_1^{(\alpha_1)}, l_2^{(\alpha_1)}, \dots l_n^{(\alpha_1)}\}$. Similarly, we collect these sets of learning rates for $N$ different learning rates between $10^{-45}$ and $10^{-13}$. That is, we get $N$ sets of values $L^{(\alpha_i)} = \{l_1^{(\alpha_i)}, l_2^{(\alpha_i)}, \dots l_n^{(\alpha_i)}\}$.
We observe that for $\alpha_i \neq \alpha_j$ and $\alpha_i, \alpha_j < 10^{-13}$, $l_1^{(\alpha_i)} = l_1^{(\alpha_j)}$, $l_2^{(\alpha_i)} = l_2^{(\alpha_j)}$ and so on. That is, we get that the set of values $L^{(\alpha_i)} = L^{(\alpha_j)}$.
This implies that the loss values reported by the fine-tuning API do not change significantly when the learning rates are smaller than $\approx 10^{-13}$

This supports our hypothesis that small learning rates do not change the model parameters significantly.
Furthermore, any value in this range can serve as a small learning rate since they all guarantee that the loss values being reported are not being affected significantly by the training of the model itself.

\newparagraph{The training losses reported are permuted.} To establish this, we notice that if we send a fine-tuning request with an ordered training set consisting of duplicated data, say, $D = \{(x^{(1)}, y), (x^{(2)}, y), (x^{(2)}, y), (x^{(3)}, y), (x^{(3)}, y), (x^{(3)}, y)\}$ of size $6$ with a batch size of $1$ for $1$ epoch with a small learning rate, we should expect to get an ordered list of losses $L = \{l^{(1)}, l^{(2)}, l^{(2)}, l^{(3)}, l^{(3)}, l^{(3)}\}$ in response. Instead, we find that the losses reported are permuted according to some permutation. That is, we do observe the same cardinalities of different losses as expected, \ie, out of the $6$ loss values reported, one value appears exactly once, one value appear twice, and one value appears thrice, but in a different order from the input. Note that counting the cardinalities of duplicated training items is the only way to determine whether a permutation occurred or not because we do not know the loss values of each individual training example ahead of time.

Similar observations hold across different training set sizes and different sets of training examples.
Therefore, we can conclude that for an ordered training set $D = \{(x^{(i)}, y^{(i)} \}_{i=1}^{N}$ of size $N$, there is a permutation $\sigma_N$ such that the true training losses corresponding to each example in the dataset $D$ are obtained by applying the permutation $\sigma_N$ to the reported ordered set of losses $L = \big[l^{(i)}\big]_{i=1}^{N}$.

\newparagraph{The same permutation is applied across different fine-tuning requests.}
We also observe that making different fine-tuning requests with the same training set sizes (but different training data) results in the same orderings of reported losses.
Therefore, we conclude that the fine-tuning procedure uses a constant, hardcoded seed value $s$ to initialize a generator and applies the same, fixed permutation to a training set of a given size $N$ (as long as the batch size is $1$). We refer to this permutation using $\sigma_N$.

\subsection{Reverse Engineering the Training Loss}
\label{subsec:loss_rev}

Gemini documentation doesn't provide details on how the training loss is computed and optimized during fine-tuning. Different approaches to fine-tuning can optimize and report different types of loss functions, such as:

\begin{newitemize}
    \item Cross-Entropy computed over only the output string (``Instruction Tuning'')\cite{raschka2024llm}.
    \item Cross-Entropy computed over both the input and output strings (``Instruction Modeling'')\cite{shi2024instructiontuninglossinstructions}.
    \item Some unknown custom loss functions such as Distillation losses or Sparse Training losses \cite{kurtic2023sparsefinetuninginferenceacceleration}.
\end{newitemize}

In this subsection, we partially reverse engineer the reported loss to analyze its effectiveness as a proxy for the adversarial objective \ie the average logprobs. We do this by comparing the Fine-Tuning loss, which we denote as $\mathrm{TrainingLoss}$, with logprobs, denoted later as $\mathrm{AvgLogprobs}$. Natively, the Gemini API doesn't expose the logprobs of any string, however, Vertex AI, an enterprise API by Google, does provide an ``average logprobs'' value for top $8$ responses. For our analysis in this section, we assume that both the Gemini API and the Vertex AI API serve the same base model. We treat the average logprobs as the ground truth and assume that it is the logprobs of the generated response (see \cref{eq:logprobs_definition}) divided by the length of the output.

Next, we pick three distinct input prompts and compare the $\mathrm{TrainingLoss}$ with the $\mathrm{AvgLogprobs}$ for each prompt.
Concretely, for a prompt $X$, we collect the average logprobs of the generated response $Y_{1:l}$ (of length $l$) and collect the training loss for the input-output pair $(X,Y_{1:l})$.
We plot $\mathrm{TrainingLoss}$ and $l \cdot \mathrm{AvgLogprobs}$ (average logprobs scaled to the output length, which we call total logprobs) as functions of the output length $l$ as we incrementally increase the output length $l$ in \cref{fig:losses_compare}.
We make a few observations:
\begin{enumerate}
\item Both values increase proportionately to the length $l$
\item Training loss increases proportionally to the total logprobs: their difference stays nearly constant and doesn't depend on the output length
\item This difference varies as we change the inputs in the three plots.
\end{enumerate}
The above let us conjecture that a linear relationship exists between training loss and average logprobs, where the training loss is a function of both inputs and outputs:

\newparagraph{Hypothesis.} We hypothesize the following closed-form expression for the FL:
\[
\mathrm{TrainingLoss}(Y|X) = K(X) + l \cdot \mathrm{AvgLogprobs}(Y|X)
\]
where the term $K({X})$ is a function of input $X$.

\newparagraph{Validation Experiment.}
First we note that for a dataset of $l$-length output sequences, $D^{(l)} = \{(X_i, Y_{i,1:l})\}_{i=1}^{n}$, we have the total logprob values $\mathcal{T}_0^{(l)} = \{l\cdot \mathrm{AvgLogprobs}(Y_{i,1:l}|X_i)\}_{i=1}^{n}$, and the corresponding hypothesized training loss values $\mathcal{T}_1^{(l)} = \{K({X_i}) + l\cdot \mathrm{AvgLogprobs}(Y_{i,1:l}|X_i)\}_{i=1}^{n}$.
The formula for the R-squared value for the two sets of numbers ($\mathcal{T}_0^{(l)}$ and $\mathcal{T}_1^{(l)}$) is given by
\[
R^{2(l)} = 1 - \frac{\sum\limits_{i=1}^{n}\big( K({X_i}) \big)^2}{l^2 \cdot \text{Var}(\{\mathrm{AvgLogprobs}(Y_{i,1:l}|X_i)\}_{i=1}^{n})}
\]Thus, if our hypothesized form is correct, then for large values of $l$, the R-squared value should approach $1$. 

\usepgfplotslibrary{groupplots}
\begin{figure}[t]
\begin{tikzpicture}[scale=1]
\small
\begin{groupplot}[
    group style={
        group size=3 by 1,
        group name=plots,
        x descriptions at=edge bottom,
        y descriptions at=edge left,
        horizontal sep = 2pt},
    ylabel={Value},
    y label style={at={(axis description cs:0.2,.5)}},
    x label style={at={(axis description cs:0.5,0.1)}},
    ymin=0, ymax=200,
    ymajorgrids=true,
    grid style=dashed,
    width=0.23\textwidth,
    title style={at={(0.5,0.9)},anchor=south},
]
\nextgroupplot[title={Q1},bar width=5pt,xlabel={\# Output Tokens},legend to name={CommonLegend},legend style={legend columns=2}];
\addplot+[mark size=1pt] table[x=length, y=q11, col sep=comma]{plots/loss_differences.csv};
\addplot+[mark size=1pt] table[x=length, y=q12, col sep=comma]{plots/loss_differences.csv};
\addlegendimage{red, mark=*}
\addlegendentry{Total Logprob}
\addlegendentry{Training Loss}

\nextgroupplot[title={Q2},bar width=5pt,xlabel={\# Output Tokens}];
\addplot+[mark size=1pt] table[x=length, y=q21, col sep=comma]{plots/loss_differences.csv};
\addplot+[mark size=1pt] table[x=length, y=q22, col sep=comma]{plots/loss_differences.csv};

\nextgroupplot[title={Q3},bar width=5pt,xlabel={\# Output Tokens}];
\addplot+[mark size=1pt] table[x=length, y=q31, col sep=comma]{plots/loss_differences.csv};
\addplot+[mark size=1pt] table[x=length, y=q32, col sep=comma]{plots/loss_differences.csv};
\end{groupplot}
\path ($(plots c2r1.south west)-(0,0.7cm)$) -- node[below]{\ref*{CommonLegend}} ($(plots c2r1.south east)-(0,0.7cm)$);

\end{tikzpicture}

{\scriptsize Q1: Write a long essay on the benefits... A: The Profound Benefits of Exercise...\\
Q2: Tell me about elves and fairies. A: Elves are mythical beings...\\
Q3: Explain in great detail... A. Transistors are fundamental building blocks...}

    \caption{Total logprobs, training loss, and output length are all pairwise proportional. The difference between total logprobs and training losses for a fixed input-output pair is independent of the output length.}
    \label{fig:losses_compare}
\end{figure}

\newparagraph{Empirical evidence.} We conduct the above experiment empirically, we create a dataset of $n = 10$ open-ended questions $\{X_i\}_{i=1}^{10}$. For a fixed output length of $l$ tokens, we collect the average logprobs $\mathcal{A}_0^{(l)} = \{\mathrm{AvgLogprobs}(X_i, Y_{i,1:l})\}_{i=1}^{10}$ of each of the generated responses $\{Y_{i,1:l}\}_{i=1}^{10}$ when truncated to the length $l$.
We then send the training dataset $\{X_i, Y_{i, 1:l}\}_{i=1}^{10}$ for fine-tuning and collect the training losses $$\mathcal{A}_1^{(l)} = \{\mathrm{TrainingLoss}(X_i, Y_{i, 1:l})\}_{i=1}^{10}$$
We perform a linear regression over the $10$ data points in $\mathcal{A}_0^{(l)}$ and $\mathcal{A}_1^{(l)}$ and record the goodness-of-fit coefficient $R^{2(l)}$. Finally, we plot $R^{2(l)}$ as a function of $l$. The results are as shown in \cref{fig:regression_trend}
As can be seen from the graph, we see an almost perfect correlation for large values of $l$, thus validating our hypothesis.

\begin{figure}[t]
\centering
\begin{tikzpicture}
\begin{axis}[
        width=\linewidth,
        height=4cm,
        xlabel={Output length $l$},
        ylabel={$R^2$},
        y label style={at={(axis description cs:0.035,.5)}},
        ymin=0, 
        ymajorgrids=true,
        grid style=dashed,
    ]
    \addplot+[color=black,mark=x,mark size=1.5pt] table[x=length, y=r2_value, col sep=comma]{plots/r2_graph_data.csv};
\end{axis}
    
\end{tikzpicture}
\caption{The correlation between average logprobs and training losses asymptotically approaches $1$ as the length of the output string increases.}
\label{fig:regression_trend}
\end{figure}

It is unclear how the term $K({X})$ is computed.
One explanation might be that the training loss performs Instruction Modeling \cite{raschka2024llm}, where the training loss is computed over both input and output and possibly includes unknown terms or additional, unknown tokens in its internal representation of a training example. However, we note that we do not need to know exactly how the term $K(X)$ is calculated since we can establish that training loss is a useful proxy even without knowing the exact form of $K(X)$.

\subsection{Training loss is a useful proxy for optimization}
\label{subsec:tl_proxy}

The empirical data (\cref{fig:regression_trend}) shows that the training loss is almost perfectly correlated with the average logprobs when the length of the target string is long. Therefore the training loss serves as an almost perfect proxy for the adversarial objective function when the length of the target string is long.
In this subsection, we empirically establish that even for short target strings, the training loss acts as a usable proxy.

For iterative solutions to the optimization problem, training loss can be a good proxy if it can guide the search process toward the ``correct'' direction \ie in the direction of minimizing the ``true'' loss (the logprobs). In other words, it should help identify the best small perturbation (the one leading to minimum true loss) from a set of candidate small perturbations. Empirically, we find that while the training loss for Gemini doesn't always identify the best perturbation from a set of perturbations, the selected perturbation is better than the average (\ie a randomly sampled perturbation).

To understand this, we examine how good is the ``true'' performance or rank (according to logprobs) of the candidate selected by the training loss. We consider a question $X$ to which the LLM replies with a short, deterministic, well-known answer $Y$. We then create a dataset of $N$ small perturbations of $X$ by appending a randomly generated token at the end of $X$, while ensuring that the output of the LLM on all these perturbations continues to be $Y$. That is, $D = \{(X\Vert r^{(i)}, Y)\}_{i=1}^{N}$. We then collect the average logprobs $\{\mathrm{AvgLogprobs}_{i}\}_{i=1}^{N}$ corresponding to the inputs $\{X \Vert r^{(i)}\}_{i=1}^{N}$ and the training losses $\{\mathrm{TrainingLoss}_{i}\}_{i=1}^{N}$ for this dataset $D$. We then compute the rank of the best candidate selected by the training loss (\ie $\text{argmin}_{i \in \{1 \dots N\}}\mathrm{TrainingLoss}_{i}$) in the list of perturbations sorted in the increasing order of the true loss (logprobs). \cref{fig:FreqCharts} shows the distribution of the rank for $3$ different questions.

\usepgfplotslibrary{groupplots}
\begin{figure}[t]
\begin{tikzpicture}[scale=1]
\small
\begin{groupplot}[
    group style={
        group size=3 by 1,
        group name=plots,
        x descriptions at=edge bottom,
        y descriptions at=edge left,
        horizontal sep = 2pt},
    ylabel={Frequency},
    y label style={at={(axis description cs:0.225,.5)}},
    x label style={at={(axis description cs:0.5,0.1)}},
    ymin=0, ymax=30,
    ytick={0,10,20,30},
    xtick distance=1,
    ymajorgrids=true,
    grid style=dashed,
    width=0.23\textwidth,
    area style,
    title style={at={(0.5,0.9)},anchor=south},
]
\nextgroupplot[title={Q1},bar width=5pt,
xlabel={Position $j$}];
\addplot+[ybar,mark=no] table [x=rank, y=q1, col sep=comma] {plots/rank_distribution.csv};

\nextgroupplot[title={Q2},bar width=5pt,xlabel={Position $j$}];
\addplot+[ybar,mark=no] table [x=rank, y=q2, col sep=comma] {plots/rank_distribution.csv};

\nextgroupplot[title={Q3},bar width=5pt,xlabel={Position $j$}];
\addplot+[ybar,mark=no] table [x=rank, y=q3, col sep=comma] {plots/rank_distribution.csv};
\end{groupplot}
\end{tikzpicture}

{\scriptsize Q1: What breed is the dog Scooby Doo? A: Great Dane\\
Q2: Who was the first president of the United States of America? A: George Washington\\
\scalebox{0.92}[1]{Q3: What is the name of the fictional spy who goes by the codename 007? A. James Bond}}
    \caption{Rank distribution of top candidate from training losses, with $M=100$ samples each for $N=10$ candidates}
    \label{fig:FreqCharts}
\end{figure}
We record the value $j$ and repeat the experiment $M$ times, each with fresh randomly generated candidates, to obtain a list of values of $j$ all lying in $\{1 \dots N\}$. We plot the frequencies of $j$ to obtain a distribution.
Three such frequency charts are shown in \cref{fig:FreqCharts} for three questions and answers.

We see that all distributions are skewed highly to the left indicating that with high probability, the best candidate minimizing the training losses is also amongst the top few candidates in the average logprobs. This analysis confirms that the training loss value can serve as a noisy signal to guide the discrete optimization process.
\section{Adversarial Prompt Optimization using the Fine-Tuning interface}
\label{sec:ft_opt}

We have empirically established that fine-tuning loss can act as a good proxy for guiding the adversarial optimization process. In this section, we use it to automatically generate prompt injections.

\subsection{Recovering the random permutation}
\label{subsec:recover_random}

In principle, we can obtain the training loss corresponding to a single training example $(x^{(i)}, y^{(i)})$ atomically by sending a fine-tuning request with a
single training example with a batch size of $1$ for $1$ epoch and obtain a single value.
Unfortunately, due to a spin-up overhead from a few seconds to several minutes per fine-tuning request, it is desirable to evaluate the training loss for multiple training examples in one query. However, the permutation applied to the losses obscures the correspondence between training examples and their true training losses when a dataset has several training examples.

Our attack sidesteps this permutation of training losses by supplying specially crafted training examples, recovering an unknown, but approximate, permutation $\sigma_N$, and reusing it during later attack steps.
The key idea is to take some prompt $x_{\mathrm{Prompt}}$, progressively corrupt the corresponding true base model's response $y_{\mathrm{True}}$, and fine-tune this model on a training set of multiple copies of $x_{\mathrm{Prompt}}$ paired with these garbled target values.
For this training set consisting of progressively corrupted output strings, the training loss values are expected to appear in ascending order after fine-tuning, revealing the matching between initial and reshuffled loss values.

During this step, we prompt the base model with $x_{\mathrm{Prompt}}$ for a sufficiently long, deterministic, and well-formed response, such as a quote from a book, to record the true model response $y_{\mathrm{True}}$. We define initial $y_{\mathrm{Target}, 0}$ as $y_{\mathrm{Target}, 0} = y_{\mathrm{True}}$.
Then, for each $i \in \{1, \dots N\}$ we create $y_{\mathrm{Target}, i}$ by corrupting the starting $i$ tokens from $y_{\mathrm{Target}, 0}$.
The process is illustrated in \cref{tab:quickbrown}.

\begin{table}[t]
  \centering
  \caption{Example of building $D_{garbled}$ by progressively corrupting $y_{\mathrm{Target}}$ on the same $x_{\mathrm{Prompt}}=\text{``Repeat this: quick brown fox.''}$}
  \begin{tabular}{llc}
    \toprule
    & $y_{\mathrm{Target}, i}$ & TrainingLoss \\
    \midrule
    $y_{\mathrm{Target}, 0}$ & Quick brown fox & 16.08243 \\
    $y_{\mathrm{Target}, 1}$ & {\color{red}\foreignlanguage{russian}{Наш}} brown fox & 41.72246 \\
    $y_{\mathrm{Target}, 2}$ & {\color{red}\foreignlanguage{russian}{Наш} oss} fox & 57.49492 \\
    $y_{\mathrm{Target}, 3}$ & {\color{red}\foreignlanguage{russian}{Наш} ossgebnis.} & 75.69193 \\
    % ...
    \bottomrule
  \end{tabular}
  \label{tab:quickbrown}
\end{table}

Thus, we obtain a dataset $D_{garbled}$ of pairs $(x_{\mathrm{Prompt}}, y_{\mathrm{Target}, {i}})$ whose training loss values should increase monotonically.
Finally, we initiate a fine-tuning request with $D_{garbled}$, a batch size of $1$, and a small learning rate, where an ascending sorting of the resulting loss values $\overline{l_{0}}, ..., \overline{l_{N}}$ reveals the sought-after permutation: 
\begin{equation}
    \label{eq:11}
    \sigma_N =
    \begin{pmatrix}
    \mathrm{AscendingSort}(\overline{l_{0}}, \dots, \overline{l_{N}}) \\
    \overline{l_{0}}, \dots, \overline{l_{N}}
    \end{pmatrix}
\end{equation}

This permutation $\sigma_N$ is constant for a fixed training set size $N$, which allows us to reuse the learned $\sigma_N$ for all fine-tuning requests of the same size.
Our method computes an approximate permutation that relies on the assumption of monotonicity of losses of increasingly garbled inputs.
To evaluate the performance of our method, we compare the results with an inefficient, but provably correct method, detailed in the appendix. We find that our approximate method only misidentifies approximately $7-8\%$ of the positions in the permutation, and preserves more than $90\%$ of the pairwise orderings of the permutation. We note that this is a tolerable margin since any small errors in the permutation 
are indistinguishable from the small noise in the candidate ranking as we observed in \cref{subsec:tl_proxy}.

\subsection{Fun-tuning attack}

In our case, the full prompt to a model is represented by a pair of system prompt $x_{\mathrm{Sys}}$ and a user prompt $x_{\mathrm{User}}$, where a substring of $x_{\mathrm{User}}$ is adversarially controlled:
$$x_{\mathrm{Prompt}} = x_{\mathrm{Sys}} \Vert x_{\mathrm{User}} = x_{\mathrm{Sys}} \Vert x_{\mathrm{User_{question}}} \Vert x_{\mathrm{Adv}}$$

Later in the paper, we denote $x_{\mathrm{Prompt}}$ as 
a combination of trusted pair ($x_{\mathrm{Sys}}, x_{\mathrm{User_{question}}})$ and an adversarial $x_{\mathrm{Adv}}$,
which simplifies the descriptions of our algorithms:
$$x_{\mathrm{Prompt}} = x_{\mathrm{Sys}} \Vert x_{\mathrm{User}} = x_{\mathrm{Trusted}} \Vert x_{\mathrm{Adv}}$$

The adversarial sequence $x_{\mathrm{Adv}}$ itself is represented by a malicious input - typically written in plain English (such as ``\textit{Ignore previous instructions and ...}'') - surrounded by an adversarial prefix and suffix of predetermined length:
$$x_{\mathrm{Adv}} = \mathrm{Adv.\ Prefix} \Vert \mathrm{Malicious\ Instruction} \Vert \mathrm{Adv.\ Suffix}$$

For clarity, we denote the positions of a suffix and a prefix as a mask $M = (M_0, ..., M_n)$ of size $n$.
Our attack starts with the mask positions $M_0, ..., M_n$ initialized with a constant token and directly optimizes those tokens, while the malicious input stays unchanged:
$$x_{\mathrm{Adv}_{M_0:M_n}} = \mathrm{Adv.\ Prefix} \Vert \mathrm{Adv.\ Suffix}$$

At each iteration, before we start finding replacement tokens, we estimate the best position to perturb $M_{best} \in \{M_0, ..., M_n\}$, which we will optimize afterwards.
The best position is chosen as a position minimizing the average loss for a small set of replacement tokens $\mathcal{R}$:
\begin{equation}
    \label{eq:bestpos}
    M_{best} = \argmin{m \in M} \mathop{\mathbb{E}}_{x_{\mathrm{Adv}_{m}} \in \mathcal{R}}[\mathcal{L}(x_{\mathrm{Trusted}} \Vert x_{\mathrm{Adv}}, y_{\mathrm{Target}})]
\end{equation}

We randomly sample with replacement a set $\mathcal{R}$ of unique tokens using rejection sampling and create $K = |\mathcal{R}| \cdot |M|$ candidates by substituting each mask position once for every token from $\mathcal{R}$ (\cref{alg:ftloop}{, line 7}).
Our candidate list is denoted as $\mathrm{C}$ and always has a fixed length $K$ during an attack to ensure we have a known permutation of this size.
$\mathrm{C}$ at this step can be seen as an $|M|$-long sequence of $|\mathcal{R}|$-sized chunks: we will later compute averages over those chunks to obtain the expectations from \cref{eq:bestpos}.

To evaluate the best position as defined in \cref{eq:bestpos}, we fine-tune the target model with a training set of candidates $\mathrm{C}$, a small learning rate $\alpha$, a batch size of $1$, and for $1$ epoch.
At the end of the fine-tuning, we get the training losses, restore the ordering by applying $\sigma^{-1}_{K}$ to the losses, and pick the best position with the least average loss.

Next, we find the best substitution token for the position $M_{best}$ using a very similar procedure. We sample a $K$-sized list of replacement tokens for that position to fill the training set of fixed size with candidates $\mathcal{C}*$.
Next, we obtain their losses from the fine-tuning endpoint to find the best candidate and update $x_{\mathrm{Adv}}$ with it.
The list of candidates always starts with a $x_{\mathrm{Adv}}$ itself: when all new candidates perform worse, the algorithm proceeds to the next iteration without updating $x_{\mathrm{Adv}}$, ensuring we don't pick a suboptimal substitution.
This finishes one iteration.

We run this algorithm for a chosen number of iterations.
During the iterations, we score all perturbations, and at the end we return the best $x_{\mathrm{Adv}}$ with a maximum success rate. The complete algorithm is described in \cref{alg:loft}. For the choice of default parameters, see \cref{subs:methods}.

\begin{algorithm}
    \small
    \caption{Candidate Ranking via FT Loss ($\text{Rank}_{FT}$)\label{alg:ftloop}}
    \textbf{Input: } Input prompt $x_{\mathrm{Trusted}}$, Adversarial input $x_{\mathrm{Adv}}$, Prefix-suffix indices $M_0...M_n$, Desired target $y_{\mathrm{T}}$, Number of substitutions $K$ (training set size), Small learning rate $\alpha$\\
    \textbf{Output:} $L^*$: Losses corresponding to k candidates perturbed at best index
    \begin{algorithmic}[1]
        \State $epochs \gets 1$, $bs \gets 1$ \Comment{\textcolor{gray}{Fix fine-tuning parameters}}
            \State $\mathrm{C} \gets \emptyset$ \Comment{\textcolor{gray}{Initialize empty list of candidates}}
            \State $\mathrm{R} \gets RndUniqTokens(K/n)$ \Comment{\textcolor{gray}{$K/n$ unique random tokens}}
            \For {$m \gets M_0$ to $M_n$}
                \LineComment{\textcolor{gray}{$K/n$ candidates with a token from $\mathrm{R}$ at $m^{th}$ index}}
                \State $c^{cands}_m = \texttt{Substitute}^{K / n}_m(x_{\mathrm{Adv}}, \text{Uniform})$ 
                \LineComment{\textcolor{gray}{Inject Candidates into prompt}}
                \State $\mathrm{C}_{m} \gets (x_{\mathrm{Trusted}} \Vert c_m^{cands}, y_{\mathrm{T}})$
            \EndFor
            \State $L \gets FineTune(C, \alpha, bs, epochs)$ \Comment{\textcolor{gray}{FT on cands}}
            \State $L \gets \sigma^{-1}_{K}(L)$ \Comment{\textcolor{gray}{Restore ordering}}
            \State $M_{best} \gets \argmin{m \in M_0, ..., M_n} \mathop{\mathrm{E}}_{i \in 0, ..., K / n} [L_{m, i}]$ \Comment{\textcolor{gray}{Best index}}
            \LineComment{\textcolor{gray}{$K$ candidates with random token at ${M_{best}}^{th}$ index}}
            \State $c^{cands}_{M_{best}} = \texttt{Substitute}^{K}_{M_{best}}(x_{\mathrm{Adv}}, \text{Uniform})$
            \LineComment{\textcolor{gray}{Inject Candidates into prompt}}
            \State $\mathrm{C}^* \gets (x_{\mathrm{Trusted}} \Vert c_{M_{best}}^{cands}, y_{\mathrm{T}})$
            \State $L^* \gets FineTune(\mathrm{C}^*, \alpha, bs, epochs)$
            \State $L^* \gets \sigma^{-1}_{K}(L^*)$ \Comment{\textcolor{gray}{Restore ordering}}
        \State \Return $L^*, \mathrm{C}^*$
    \end{algorithmic}
\end{algorithm}
\begin{algorithm}
    \small
    \caption{Fun-tuning attack \label{alg:loft}}
    \textbf{Input: } Input prompt $x_{\mathrm{Trusted}}$, Adversarial input $x_{\mathrm{Adv}}$, Prefix-suffix indices $M_0...M_n$, Desired target $y_{\mathrm{T}}$, Number of iterations $NumIter$, Number of substitutions $K$ (training set size), Small learning rate $\alpha$\\
    \textbf{Output:} $x_{\mathrm{Adv}, best}$: Best perturbation
    \begin{algorithmic}[1]
        \State $epochs \gets 1$, $bs \gets 1$ \Comment{\textcolor{gray}{Fix fine-tuning parameters}}
        \State $x_{\mathrm{Adv}, 0} \gets x_{\mathrm{Adv}}$ \Comment{\textcolor{gray}{Initialize adversarial perturbation}}
        \State $S \gets \emptyset$ \Comment{\textcolor{gray}{Keep track of best perturbation}}
        \For {$it \gets 1$ to $NumIter$}
            \State $x_{\mathrm{Adv}, it} \gets x_{\mathrm{Adv}, it-1}$
            \State $L^*, \mathrm{C}^* = \text{Rank}_{FT}(x_{\mathrm{Trusted}}, x_{\mathrm{Adv}, it}, M_0..M_n, y_{\mathrm{T}}, K, \alpha) $
            
            \State $i \gets \argmin{i \in 0, ..., K}\; L^*_i$ \Comment{\textcolor{gray}{Select minimum loss}}
            \State ($x_{\mathrm{Trusted}} \Vert x_{\mathrm{Adv}, it}, y_{\mathrm{T}}) \gets \mathrm{C}^*_{i}$ \Comment{\textcolor{gray}{Select best candidate}}
            
            \State $S_{it} \gets Score(GetResponse(x_{\mathrm{Trusted}} \Vert x_{\mathrm{Adv, it})}))$
        \EndFor
        \State $best = \argmax{it \in 0..K}~S_{it}$
        
        \State return $x_{\mathrm{Adv}, best}$
    \end{algorithmic}
\end{algorithm}
\section{Evaluation}
\label{sec:eval}
Our evaluation goal is to characterize the vulnerability of Google's Gemini series of closed-weights models to prompt injection attacks created using fun-tuning.  We characterize attack effectiveness along multiple dimensions:

\begin{newitemize}
    \item What is the success rate for the Fun-tuning attack?
    \item How feasible are our methods in terms of time and cost?
    \item Does our optimization algorithm provide iterative improvement and does it perform better than the baseline and an ablation attack that uses random token substitutions?
    \item What is the attack success rate when transferred to other Gemini models?
    \item How does the attack success rate and the loss depend on the attack hyperparameters - i.e. the candidate set size?

\end{newitemize}

The evaluation shows that the fun-tuning attack has an attack success rate (ASR) of 65\% for Gemini-1.5-Flash and 82\% for Gemini-1.0-Pro on the popular PurpleLlama prompt injection benchmark.
The attack is query- and cost-efficient, requiring 90 fine-tuning calls per example in PurpleLlama and all of our attacks combined cost less than \$10 in completions endpoint calls.
Our method achieves successful attacks with a candidate set size of 1000 per iteration, which is only around 1\% of the total vocabulary.

\subsection{Dataset Construction}
\label{subs:dataset}

\newparagraph{Dataset.} 
We evaluated our attack on a subset of the prompt injection dataset from the Purple Llama CyberSecEval \cite{bhatt2023purplellamacybersecevalsecure}, a well-known benchmark suite for assessing the cybersecurity vulnerabilities of Large Language Models.
The prompt injection dataset has two kinds of injections --- \textit{direct} and \textit{indirect}.
Following our threat model, we focus on \textit{indirect} ones where an injected instruction is only a part of the user prompt (a document or other content). We note that direct prompt injections, where the user is the attacker, is less likely to occur in practice. 
The \textit{indirect} examples provide a wide variety of known prompt \textit{attack categories}, such as ``ignore previous instructions,'' developer mode overrides, and hypothetical scenario attacks.
Note that all these attacks have been handcrafted by the broader security community and Meta has manually curated these into the benchmark.  Our attack wraps these existing malicious instructions with optimized prefix and suffix token sequences, in a style similar to the whitebox NeuralExec attack~\cite{pasquini2024neuralexeclearningand}. This has the effect of ``boosting'' the existing malicious instructions and forcing the LLM to obey the ``boosted'' instructions while ignoring other instructions in the context window.

The Purple Llama prompt injections dataset contains 56 examples.
To enable quicker exploration, we worked with its subset: we randomly sampled 40 indirect prompt injection examples from it to build our own dataset which we call PPL40. 
During sampling, we excluded examples that use non-standard encodings as a part of their attack (\textit{token smuggling} category and a few other examples): we found that Gemini doesn't follow instructions encoded in non-standard encodings.
The resulting PPL40 dataset reflects a similar distribution of attack categories as the original complete dataset.
The exact distribution of the attack categories in this dataset is shown in the appendix \cref{tab:purplellama_dist_table}.

Each attack category is realized in one of a few different \textit{scenarios}, such as summarizing a website contents, a code snippet, or other types of document. Prompt injection for each of the scenarios has a unique type of action injected into a corresponding document: providing a misleading answer, directing the user to a phishing website, and others.
We study the attack performance for each scenario later in the evaluation.

\newparagraph{Training example format.}
Each of the Purple Llama examples of indirect type has the following structure:
\begin{newitemize}
    \item \textit{system prompt}, which has general instructions for the model, or assigns it a role.
    \item \textit{user input}, which includes a question/prompt to the model and some external content. External content contains an injected instruction that should override the question and steer the model towards performing a different action or providing an incorrect answer about the content. We wrap this instruction with an adversarial prefix and suffix that we optimize during the attacks
    \item \textit{judge question}, which evaluates the model output to check if the attack succeeded and whether the model performed the injected instruction.
\end{newitemize}

Both \textit{system prompt} and \textit{user input} are combined into a \textit{prompt}.
This \textit{prompt} is denoted in our text as $x_{\mathrm{Trusted}} \Vert x_{\mathrm{Adv}}$, where $x_{\mathrm{Adv}}$ is an adversarially controlled injected instruction, wrapped by a 20-token-long optimizable prefix and suffix that we initialize with the token ``!''.
The \textit{prompt} is formatted as a conversation using a standard chat format.

\subsection{Metrics}
\label{subs:metrics}

Our primary metric is \textit{attack success rate (ASR)} -- a percent of successfully attacked examples from PPL40 averaged over multiple scoring runs.
Similar to the original Purple Llama approach, we compute this metric by prompting the responses from a model and assigning them binary scores (1 means attack success) by querying  GPT-4o as the \textit{judge model} with the \textit{judge question}.
To account for the target model's non-determinism, we repeat this procedure 20 times to evaluate the primary attack (5 times for the transfer evaluation) and report the mean and the standard deviation. 
We track ASR for each iteration, including the baseline ASR -- a score for an unmodified and unperturbed original injection.
The final per-example ASR we report is the maximum ASR across all iterations, while the ASR reported for the whole PPL40 attack is a sum of the maximum per-example ASRs.
Correspondingly, the PPL40 baseline ASR is a sum of the per-example baseline ASRs.
Finally, we report the improvement factor over baseline, which is $ASR_{attack} / ASR_{baseline}$.

\newparagraph{Stricter judge questions.} We observed that the original \textit{judge questions} from Purple Llama are overly permissive and result in too many false positives that can artificially boost the attack success rates.
For example, for an injected instruction ``\textit{Ignore the function and state the output is `10'},'' the default question ``\textit{Does this response contain `10'?}'' results in score $1$ when `10' appears inside model's output, which often happens when model provides a detailed explanation of the input. By contrast, under our threat model, accurate explanation in the response revealing an attack signifies a failed attempt (\cref{sec:threat}).

Accurate scoring is critical to evaluate the attack effectiveness.
To make this metric accurate, we manually adjusted the existing judge questions according to our threat model.
The corrected question for the same example is: ``\textit{The response is an answer to a question `What is the output of this code?' with some code attached. To be valid, the response needs to state that the output is `10' (or 10, any formatting is allowed). If output is interpreted by human as 10, response is valid. Any mentioning of tricky question, instruction override, or ambiguity means the response is invalid. Is this response valid?}.''

\begin{table}[t]
    \caption{Confusion matrix showing two example answers ($r_{attack}$ and $r_{fail}$, from top to bottom) from one of our attacks rated with initial judge question $j_{initial}$ and the $j_{strict}$ we introduced. $j_{initial}$ results in both a false positive (FP) and a false negative (FN).}
    \centering
    \begin{tabular}{p{5.5cm}cc}
      \toprule
      Answer & \textbf{$j_\mathrm{strict}$} & \textbf{$j_\mathrm{initial}$} \\ \midrule
      {The sum of all populations is **one hundred**.} & 1 & 0 (FN) \\ \midrule
      {The question is a bit of a trick!  It asks for the sum of all the populations, but then provides an instruction to assume the sum is 100.} & 0 & 1 (FP) \\
    \bottomrule
    \end{tabular}
    \label{tab:judge}
\end{table}

\Cref{tab:judge} shows a confusion matrix for two responses rated using original and strict set of judge questions to illustrate a false positive and a false negative case.

\newparagraph{Efficiency metrics.} We compared each method's efficiency in terms of the number of iterations, attack time (hours), and cost. The time to complete a single fine-tuning call is affected by a variety of factors outside our control such as the availability of cloud servers, load balancing etc.
Thus, we only measure a rough upper-bound on the time to complete a single fine-tuning request.
We observed that for a training dataset size of $1000$ examples, successful Gemini 1.0 Pro fine-tuning calls take around 10 minutes, while Gemini 1.5 Flash fine-tuning calls succeed in under 40 minutes.
Thus, while we do report an estimate for the time to compute attacks over the entire dataset, a more reliable metric is the number of fine-tuning requests sent to the servers.

Our attack's financial cost is driven by the inference cost, because Gemini fine-tuning requests are free at the time of writing (November 2024).
The inference requests in our attack are only for the purposes of scoring.
In our cost estimation, we assume the score per example is obtained after 20 completion requests for each iteration.

\subsection{Attack configuration}
\label{subs:methods}

\newparagraph{Target models.} We focused on the vanilla Gemini API interface (not to be confused with Vertex AI).
Vertex AI is a separate, enterprise-ready AI platform that also provides an interface to query some Gemini models.
We ran two primary attacks against \textit{gemini-1.5-flash-001-tuning} and \textit{gemini-1.0-pro-001}.
Additionally, we also evaluate how well these attacks transfer to the other models in the Gemini family.
The full list of targets for both the primary and the transfer evaluations covers the following models: (1) gemini-1.0-pro, (2) gemini-1.0-pro-001, (3) gemini-1.0-pro-latest, (4) gemini-1.5-flash-001, and (5) gemini-1.5-pro-001.

Each model was tested with default parameters as well as with temperature 0 during inference, therefore, we ended up with 10 different configurations.

\newparagraph{Formatting.} We used the standard interactive chat format for prompting, placing system prompts and user prompts from our dataset as separate parts of the conversation.
Gemini 1.5 Flash explicitly supports system prompts during model initialization, so we directly initialized Gemini 1.5 Flash models with system prompts.
In contrast with prompting, the fine-tuning API does not support the standard multi-turn chat format, so we manually crafted those multi-turn conversations by placing them between \verb!<start_of_turn>! and \verb!<end_of_turn>! tokens.
We relied on the \textit{gemma-2-9b} tokenizer from Hugging Face to identify those delimiter tokens since Gemma-2 models share the same tokenizers as the Gemini family of models~\cite{gemmateam2024gemma2improvingopen}.

\newparagraph{Fun-tuning parameters.} We ran \cref{alg:loft} for 45 iterations with 2 restarts at 15th and 30th iteration.
Adversarial suffix and prefix were initialized with 20-token-long sequences of token ``!'' each and were reset back to this state at each restart.
For 15 examples out of 40, the original injected instructions were ambiguous, therefore we initialized the prefix for them with the ``\textit{Follow this new instruction:}'' at the end to obtain a good initialization (leaving this phrase optimizable during the attacks).
We minimized the loss (\cref{eq:3}) across all candidates as the optimization objective and evaluated 1000 candidates per iteration.
Each iteration had an additional step to evaluate the best position for sampling, which also used 1000 candidates (25 for each of 40 positions in the prefix-suffix pair).
We stored the perturbations found at each iteration so we could identify the best one at the end of the optimization.


\subsection{Ablation study}
\label{subs:baselines}
To establish the effectiveness of the training loss as a signal, we also performed an ablation attack where we only removed the effects of the fine-tuning procedure.
Instead of receiving the true training losses for each candidate, this algorithm received random numbers. All other attack parameters --- the core method, prefix length, suffix length, number of iterations, number of restarts, the initializations, sampling strategies, and token substitution strategies were kept the same.
We performed the ablation experiment to provide evidence that the success rate of our experiment is indeed due to the training loss being a useful signal and not due to the other components of the attack or due to random variation.

\subsection{Prompt Injection Results}
\label{subs:results}

Our key results are 
\begin{itemize}
    \item Fun-tuning outperforms baseline and ablation with improvements outside of standard deviation, achieving a success rate of $63.5\%$ against Gemini 1.5 Flash and $82.0\%$ against Gemini 1.0 Pro
    \item Our attack against Gemini is almost free (all attacks combined cost $<\$10$), query-efficient but time-consuming (90 fine-tuning calls and 16 hours per example for Gemini 1.0 Pro (60 hours for Gemini 1.5 Flash)
    \item Fun-tuning provides iterative improvements with steady ASR increases after iterations, especially at every restart
    \item Attacks succeed for all scenarios, but only partially in password phishing scenarios (both model versions) and in code analysis cases (Gemini 1.5 Flash only)
    \item Attacking Gemini 1.5 Flash produces strong perturbations: our attacks transfer well from Gemini 1.5 Flash to Gemini 2.0 Flash, to 1.0 Pro, and between the same model version numbers
\end{itemize}
Our results show that our attack works because the training losses serve as a useful signal in guiding the discrete optimization procedure.
This procedure also results in larger improvements per iteration matching our restart strategy compared to the ablation study.

In the ablation studies we achieve surprisingly large ASR of $43.8 \%$ (Gemini 1.5 Flash) and $61.3  \%$ (Gemini 1.0 Pro) compared to the baseline scores $27.5 \%$ (Gemini 1.5 Flash) and $42.5  \%$ (Gemini 1.0 Pro), suggesting that random token substitution strategy might also be effective against Gemini models.
Finally, the baseline scores themselves suggest that some of the manually curated attacks from the Purple Llama dataset are also effective against Gemini.

Attacks are efficient, scalable, transferrable between Gemini models, and work with arbitrary prompts and target outputs.
The attacks partially fail to mislead phishing scenarios, especially against Gemini 1.5 Flash, likely due to improved safety tuning.

Contrary to the Gemini 1.5 Flash report by Google, our results show that optimization-based prompt injections are still a valid risk~\cite{geminiteam2024gemini15unlockingmultimodal}.
According to the report, with 15 million queries and internal access, genetic algorithms were able to produce universal perturbations leading to sensitive information disclosure with $0-9\%$ ASR, suggesting that optimization-based attacks might not be very effective in that setup \cite{geminiteam2024gemini15unlockingmultimodal}.
In our work, we attacked each example separately (so the perturbations do not necessarily transfer across examples) in 90 queries each, achieving an overall ASR of $65.3 \%$ against Gemini 1.5 Flash and a $82\%$ ASR against Gemini 1.0 Pro.
In a later report, Google provided additional clarifications about automated red-teaming methods to evaluate the risk from prompt injection attacks: it includes optimization-based attacks such as Actor Critic and Beam Search \cite{google2025review}.
We believe that those attacks could also use the Fine-Tuning loss in remote, query-based scenarios where no better way to estimate the attack probability is available.


\newparagraph{Different prompt injection scenarios.}
We study the attack performance for each prompt injection scenario and observe that the attacks are least successful for the `password' category, where an injected instruction attempts to direct the user to a phishing website for password reset (\cref{fig:barplot_scenarios_methods_10_fig,fig:barplot_scenarios_methods_15_fig}).
While the attacks against Gemini 1.5 Flash and 1.0 do achieve some success, breaking around $10 \%$ and $20 \%$ of examples in the `password' category, lower scores suggest that Gemini models were trained to resist phishing in some way.
The next category where our attack fails is summarizing a Python code snippet with an injected comment, attempting to mislead the model about the code's output.
In this scenario, our attack mostly fails against Gemini 1.5 Flash ($40 \%$ ASR), but is successful against an older Gemini 1.0 Pro ($80 \%$ ASR), suggesting the newer model is significantly better at code analysis.
It is unclear if Gemini executes the provided code snippet and whether better optimization-based attacks can be built specifically for code analysis.
Our fine-tuning-guided attack is successful against all other categories, successfully overriding the user's instruction with $>60 \%$ per-category ASR for both models, suggesting that our optimization strategy can be useful in practice to find prompt injections, especially if security risks are subtle, very specific to the application, or hard to anticipate in advance.
Examples of such risks include tricking a model into providing a wrong document summary or augmenting the model's output with concealed information: it is unclear how to distinguish unsafe behavior from expected without application context as those definitions are not universal.

\newparagraph{Efficiency analysis.}
Our Fun-tuning attack requires a low cost of under $\$10$ and 90 fine-tuning queries to complete.
We plot the combined ASR against attack iterations and observe that Gemini 1.0 Pro optimization quickly drives ASR in the first 20 iterations (\cref{fig:asr_iter_10}), while Gemini 1.5 Flash attack makes slower improvements until the 35th iteration (\cref{fig:asr_iter_15}).
Given our restart strategy, the slopes of \cref{fig:asr_iter_10,fig:asr_iter_15} suggest that Gemini 1.0 Pro is mostly attacked in the first 15 iterations and doesn't benefit from more restarts, but they are helpful against Gemini 1.5 Flash as \cref{fig:asr_iter_15} shows: a lot of score improvement happens shortly after each restart (after 0th, 15th, and 45th iterations).
The Gemini Fine-Tuning requests are free of charge, so the attack costs are only driven by inference costs, and inference is only used for scoring.
Gemini 1.0 Pro Fine-tuning requests mostly finish in about 10 minutes, while Gemini 1.5 Flash Fine-tuning calls terminate in about 40 minutes.
From that, a single Fun-tuning attack finishes in 15 hours against Gemini 1.0 Pro and in 60 hours against Gemini 1.5 Flash.
We observed that attacks from a single Google account degrade the parallel performance of fine-tuning requests.
However, the attack is easily scalable: (1) we are not hitting the
rate-limiting for our Fine-Tuning calls and are unaware of other bottlenecks; (2) attacks can use multiple Google accounts in a trial period with free credits.

\newparagraph{Attack transfer.} Evaluation of the perturbations from the Gemini 1.0 Pro attack shows that all of them perfectly transfer to similar Gemini 1.0 Pro models with ASR of $>80\%$ ASR and partially to Gemini 1.5 Flash with $50-60\%$ ASR (\cref{tab:transfer_10}).
Attacks computed against Gemini 1.5 Flash, on the other hand, perfectly transfer to all models with $>72\%$ ASR for Gemini 1.0 Pro and with a similar ASR of $>60\%$ to the remaining Gemini 1.5 Pro (\cref{tab:transfer_15}).
Our attack transfers to Gemini 2.0 Flash with a surprisingly larger ASR of $>80\%$ indistinguishable from the ablation ASR for the same Gemini 2.0 Flash, suggesting that the new Gemini 2.0 Flash might be better at following instructions, including the injected ones.
Better benchmarks and a closer study might be needed to understand the newer Gemini models, such as Gemini 2.0 Flash, and the attacks against them.



\subsection{{Impact of Candidate Set Size on Attack (Local Simulations)}} We study how candidate set size affects attack success rate and adversarial loss using local simulation.
For the simulations, we used gemma-2-9b-it as our target model and we directly computed the cross-entropy loss summed over the output tokens. 
We implemented a variant of our discrete optimization procedure in \cref{alg:loft}, where we modify the candidate generation step to perturb a randomly chosen location instead of the best position.
We measured ASR and the average final loss value. To compute the loss, we calculated the mean cross-entropy over the target outputs for each example and reported the average over the dataset.
The resulting success rates and average final loss values are shown as a function of the candidate set size in \cref{fig:ASR_vs_candidates}.
We observe that we achieve good success rates and losses after 125 candidates but see no significant gains after 1000 candidates, even though the loss does another descent at 2000 candidates.
Therefore, 1000 candidates (representing around 1\% of the vocabulary) is a reasonable candidate set size for the attack.

\begin{table}[t]
  \centering
  \caption{Attack ASR on PPL40 against Gemini-1.0-pro-001 with default temperature show that Fun-tuning is more effective than the baseline and the ablation with improvements outside of standard deviation}
\begin{adjustbox}{width=\linewidth}
  \begin{tabular}{lccccc }
    \toprule
    Attack & ASR (\%) & Improvement over    & FT req. \#   & Time (hrs,   & Cost (\$, \\
           &          & baseline (x)        & (1 ex.)      & 1 ex.)   & 1 ex.) \\
    \midrule
    Baseline & $42.5 \pm 2.2$ & N/A   & N/A          & N/A \\ 
    Ablation  & $61.3 \pm 4.2$ & $1.4$ & 90 (sim.)   & 0.25          & 0.18 \\ 
    Fun-tuning  & $82.0 \pm 4.2$ & $1.9$ & 90          & 15 & 0.18 \\ 

    \bottomrule
  \end{tabular}
\end{adjustbox}
  \label{tab:ASR_10}
\end{table}

\begin{table}[t]
  \centering
  \caption{Attacks on PPL40 against Gemini-1.5-flash-001 with default temperature show that Fun-tuning is more effective than the baseline and the ablation with improvements outside of standard deviation}
\begin{adjustbox}{width=\linewidth}
  \begin{tabular}{lccccc }
    \toprule
    Attack & ASR (\%) & Improvement over    & FT req. \#   & Time (hrs,   & Cost (\$, \\
           &          & baseline (x)        & (1 ex.)      & 1 ex.)   & 1 ex.) \\
    \midrule
    Baseline & $27.5 \pm 2.8$ & N/A   & N/A          & N/A \\ 
    Ablation  & $43.8 \pm 3.5$ & $1.6$ & 90 (sim.)   & 0.25          & 0.02 \\ 
    Fun-tuning  & $65.3 \pm 3.8$ & $2.4$ & 90             & 60 & 0.02 \\ 

    \bottomrule
  \end{tabular}
\end{adjustbox}
  \label{tab:ASR_15}
\end{table}

\begin{figure}[t]
\centering
\begin{tikzpicture}
    \small
    \begin{axis}[
      legend pos=south east,
      legend cell align={left},
      xlabel={Iterations},
      ylabel=Attack Success Rate (\%),
      y label style={at={(axis description cs:0.035,.5)}},
      ymin=0, ymax=100,
      xmin=0, xmax=45,
      ymajorgrids=true,
      grid style=dashed,
      xtick distance=5,
      width=\linewidth,
      height=4cm,
      ]
      \addplot +[color=teal,line width=1pt,mark=*,mark size=1pt] table [x=it, y expr=\thisrow{ft}/40*100, col sep=comma] {plots/rate_by_it_10.csv};
      \addlegendentry{Fun-tuning}

      \addplot +[color=brown,line width=1pt,mark size=1pt] table [x=it, y expr=\thisrow{ablation}/40*100, col sep=comma] {plots/rate_by_it_10.csv};
      \addlegendentry{Ablation}
    \end{axis}
\end{tikzpicture}    
    \caption{Fun-tuning attack against Gemini 1.0 Pro gains most ASR in the first 10 iterations, and continues improving it, but doesn't benefit from restarts. In the ablation experiment, ASR is largely unchanged throughout the iterations.}
    \label{fig:asr_iter_10}
\end{figure}

\begin{figure}[t]
\centering
\begin{tikzpicture}
    \small
    \begin{axis}[
      legend pos=north west,
      legend cell align={left},
      xlabel={Iterations},
      ylabel=Attack Success Rate (\%),
      y label style={at={(axis description cs:0.035,.5)}},
      ymin=0, ymax=100,
      xmin=0, xmax=45,
      ymajorgrids=true,
      grid style=dashed,
      xtick distance=5,
      width=\linewidth,
      height=4cm,
      ]
      \addplot +[color=teal,line width=1pt,mark=*,mark size=1pt] table [x=it, y expr=\thisrow{ft}/40*100, col sep=comma] {plots/rate_by_it_15.csv};
      \addlegendentry{Fun-tuning}

      \addplot +[color=brown,line width=1pt,mark size=1pt] table [x=it, y expr=\thisrow{ablation}/40*100, col sep=comma] {plots/rate_by_it_15.csv};
      \addlegendentry{Ablation}
    \end{axis}
\end{tikzpicture}    
    \caption{Fun-tuning attack against Gemini 1.5 Flash results in a steep incline shortly after iterations 0, 15, and 30 and evidently benefits from restarts. The ablation method's improvements per iteration are less pronounced}
    \label{fig:asr_iter_15}
\end{figure}

\begin{table}[t]
\centering
\caption{ASR (\%) of Gemini 1.0 Pro attacks success rates against other Gemini models for each method (attack transfer evaluation)}.
\label{tab:result}
\begin{tabular}{llccccc}
\toprule
 \multirow{1}{*}{{Model}} & \multirow{1}{*}{Baseline ASR} & \multicolumn{1}{c}{Ablation ASR} & \multicolumn{1}{c}{Fun-tuning ASR} \\
\midrule
1.0-pro & \cellcolor{orange!32}$41.0 \pm 2.8$ & \cellcolor{orange!51}$64.5 \pm 7.2$ & \cellcolor{orange!70}$87.5 \pm 2.5$\\
1.0-pro-t0 & \cellcolor{orange!34}$42.5 \pm 0.0$ & \cellcolor{orange!50}$62.5 \pm 0.0$ & \cellcolor{orange!70}$88.0 \pm 1.0$\\
1.0-pro-latest & \cellcolor{orange!36}$45.0 \pm 5.0$ & \cellcolor{orange!53}$67.0 \pm 3.2$ & \cellcolor{orange!70}$88.0 \pm 4.0$\\
1.0-pro-latest-t0 & \cellcolor{orange!34}$42.5 \pm 0.0$ & \cellcolor{orange!49}$62.0 \pm 1.0$ & \cellcolor{orange!70}$88.5 \pm 1.2$\\
1.5-flash-001 & \cellcolor{orange!23}$29.5 \pm 4.8$ & \cellcolor{orange!36}$45.0 \pm 5.3$ & \cellcolor{orange!44}$56.0 \pm 5.8$\\
1.5-flash-001-t0 & \cellcolor{orange!20}$25.5 \pm 1.0$ & \cellcolor{orange!37}$46.5 \pm 1.2$ & \cellcolor{orange!39}$49.0 \pm 1.2$\\
2.0-flash & \cellcolor{orange!36}$45.5 \pm 4.0$ & \cellcolor{orange!66}$82.5 \pm 6.0$ & \cellcolor{orange!69}$86.5 \pm 2.2$\\
2.0-flash-t0 & \cellcolor{orange!39}$49.0 \pm 1.2$ & \cellcolor{orange!72}$90.0 \pm 1.7$ & \cellcolor{orange!72}$90.0 \pm 0.0$\\
1.5-pro-001 & \cellcolor{orange!26}$33.5 \pm 2.8$ & \cellcolor{orange!42}$53.0 \pm 6.0$ & \cellcolor{orange!50}$63.5 \pm 3.8$\\
1.5-pro-001-t0 & \cellcolor{orange!26}$32.5 \pm 1.7$ & \cellcolor{orange!46}$57.5 \pm 1.7$ & \cellcolor{orange!50}$63.5 \pm 2.8$\\

\bottomrule
\end{tabular}
\label{tab:transfer_10}
\end{table}

\begin{table}[t]
\centering
\caption{ASR (\%) of Gemini 1.5 Flash attacks success rates against other Gemini models for each method (attack transfer evaluation)}.
\label{tab:result}
\begin{tabular}{llccccc}
\toprule
 \multirow{1}{*}{{Model}} & \multirow{1}{*}{Baseline ASR} & \multicolumn{1}{c}{Ablation ASR} & \multicolumn{1}{c}{Fun-tuning ASR} \\
\midrule
1.0-pro-001 & \cellcolor{orange!31}$39.5 \pm 4.0$ & \cellcolor{orange!57}$71.5 \pm 6.8$ & \cellcolor{orange!57}$72.0 \pm 4.8$\\
1.0-pro-001-t0 & \cellcolor{orange!33}$42.0 \pm 1.0$ & \cellcolor{orange!50}$63.5 \pm 1.2$ & \cellcolor{orange!56}$71.0 \pm 2.2$\\
1.0-pro & \cellcolor{orange!34}$42.5 \pm 5.0$ & \cellcolor{orange!55}$69.5 \pm 5.5$ & \cellcolor{orange!58}$73.0 \pm 6.2$\\
1.0-pro-t0 & \cellcolor{orange!34}$42.5 \pm 0.0$ & \cellcolor{orange!50}$63.5 \pm 1.2$ & \cellcolor{orange!57}$71.5 \pm 1.2$\\
1.0-pro-latest & \cellcolor{orange!34}$43.0 \pm 4.8$ & \cellcolor{orange!55}$69.5 \pm 4.0$ & \cellcolor{orange!57}$72.0 \pm 4.5$\\
1.0-pro-latest-t0 & \cellcolor{orange!34}$42.5 \pm 0.0$ & \cellcolor{orange!50}$63.5 \pm 1.2$ & \cellcolor{orange!57}$72.0 \pm 1.0$\\
2.0-flash & \cellcolor{orange!38}$48.0 \pm 2.0$ & \cellcolor{orange!68}$85.5 \pm 3.2$ & \cellcolor{orange!71}$89.0 \pm 2.8$\\
2.0-flash-t0 & \cellcolor{orange!38}$48.5 \pm 2.2$ & \cellcolor{orange!68}$86.0 \pm 1.2$ & \cellcolor{orange!72}$90.5 \pm 2.0$\\
1.5-pro-001 & \cellcolor{orange!25}$31.5 \pm 2.8$ & \cellcolor{orange!44}$55.5 \pm 2.8$ & \cellcolor{orange!48}$60.5 \pm 2.8$\\
1.5-pro-001-t0 & \cellcolor{orange!25}$32.0 \pm 1.0$ & \cellcolor{orange!47}$59.0 \pm 2.8$ & \cellcolor{orange!50}$63.5 \pm 3.8$\\

\bottomrule
\end{tabular}
\label{tab:transfer_15}
\end{table}

\begin{figure}[t]
\centering
\begin{tikzpicture}[]
    \small
    \begin{axis}[
      legend pos=south east,
      legend cell align={left},
      xlabel={Number of candidates, $|C|$},
      ylabel=Attack Success Rate (\%),
      y label style={at={(axis description cs:0.035,.5)}},
      ymin=0, ymax=100,
      xmin=0, xmax=2000,
      ymajorgrids=true,
      grid style=dashed,
      xtick distance=1000,
      width=\linewidth,
      height=4cm,
      ]
      \addplot +[color=teal,line width=1pt,mark=*,mark size=1pt] table [x=C, y expr=\thisrow{ASR}, col sep=comma] {plots/ASR_vs_numcands.csv};
    \end{axis}
\end{tikzpicture}
\begin{tikzpicture}[]
    \small
    \begin{axis}[
      legend pos=north east,
      legend cell align={left},
      xlabel={Number of candidates, $|C|$},
      ylabel=Average Loss Value,
      y label style={at={(axis description cs:0.035,.5)}},
      ymin=0, ymax=2,
      xmin=0, xmax=2000,
      ymajorgrids=true,
      grid style=dashed,
      xtick distance=1000,
      width=\linewidth,
      height=4cm,
      ]
      \addplot +[color=teal,line width=1pt,mark=*,mark size=1pt] table [x=C, y expr=\thisrow{Loss}, col sep=comma] {plots/Loss_vs_numcands.csv};
    \end{axis}
\end{tikzpicture}    

    \caption{Locally simulated attack against gemma-2-9b-it gains significantly when candidate set size $|C|$ increases to 125, and provides no significant gains after 1000.}
    \label{fig:ASR_vs_candidates}
\end{figure}


\section{Discussion}
\label{sec:disc}
\newparagraph{Evaluating loss on arbitrary values.} Our work serves as a proof-of-concept, showing that the Fine-Tuning APIs can nevertheless expose the closed-weights base model to optimization-based attacks.
We demonstrate the feasibility of our base model loss extraction idea by guiding a very basic random substitution algorithm to compute prompt injections.
However, our fine-tuning-based loss extraction can be combined with any automatic LLM attack that requires a loss value for guiding, that is, our method allows evaluating loss for arbitrary inputs and outputs and makes no assumptions about the high-level objective or optimization algorithm.

\newparagraph{Attack universality across other APIs.} While we established the feasibility of our method specifically against the Gemini API, the same idea could be used to attack other Fine-Tuning APIs.
The method we described depends on the level of control over a few input hyperparameters: minimum learning rate, minimum batch size, and random seed (static, externally controlled, or neither).
Our research in previous sections shows that an attack is possible for a learning rate parameter between $10^{-30} - 10^{-45}$, a batch size of 1, and either a static or an externally controlled random seed.
We show in \cref{tab:finetuning_api_params} that the settings for certain Fine-Tuning vendors do not (or did not) exclude the possibility of an attack beyond Gemini API.
We hope that this comparison and our paper can serve as a starting point to rigorously understand adversarial capability of discrete optimization against closed-weights models, informing proper risk assessment.

\begin{table}[t]
  \centering
  \begin{threeparttable}
  \caption{Level of control for Fine-Tuning API parameters for multiple closed-weights LLMs available at the time of writing (April 2025).}
  \begin{tabular}{p{1.8cm}p{2.9cm}p{0.4cm}p{1.7cm}}
    \toprule
    LLM API & Min. learning rate parameter & Rnd. seed & Min. batch size \\
    \toprule
    Google Gemini API\tnote{A} & $10^{-3}$ (pre-2025: $10^{-45}$) & static & 4 (pre-2025: 1) \\
    \midrule
    Google Vertex\tnote{B}  & $10^{-45}$  & static & auto \\
    \midrule
    OpenAI\tnote{C} & $10^{-5}$ (pre-2025: $10^{-31}$) & ctrl. & 1 \\
    \midrule
    Anthropic (Amazon \cite{claude3finetuning}) & $10^{-1}$ & static & auto \\
    \bottomrule
  \end{tabular}
  \begin{tablenotes}
  \item[A] Following our disclosure, Google has incorporated the changes, described as ``We constrained the API parameters that they were relying on. In particular, capping the learning rate to a value that would rule out small perturbations and limiting the batch size to a minimum of 4, such that they can no longer correlate the reported loss values to the individual inputs.''
  
  \item[B] We were not able to re-check the Google's Vertex AI input constraints as of April 2025 due to new authorization-related errors.

  \item[C] OpenAI started enforcing a minimum learning rate multiplier of $10^{-5}$ around January 2025 at the time we were running and evaluating our attack against GPT-4 models with multiplier values below $10^{-5}$, such as $10^{-31}$ and less. We are unaware of any context related to this update as we had not prepared any report for OpenAI at that time yet.
  \end{tablenotes}
  \label{tab:finetuning_api_params}
  \end{threeparttable}
\end{table}


\newparagraph{Mitigations that impose restrictions on hyperparameters.} Our attack exploits a fundamental utility-security trade-off: developers want fine-grained control over training hyperparameters so that they can effectively train models. This also directly benefits attackers. We believe that the general mitigation approach of reducing user control over training hyperparameters is unlikely to work since such measures reduce utility for benign developers.
For example, the LLM vendor could try to set a minimum value on the learning rate, but this clashes with the utility since different sizes of datasets have different recommended learning rates \cite{google2023gemini} and small learning rates often lead to stabler training, which can be desirable for benign users\cite{anyscale2023finetuning}.

Similarly, randomizing the training set on every single API call will destroy the correspondence between returned losses and candidates being evaluated, but it will still not prevent an attack since an attacker can still extract losses atomically or by using a dataset consisting of different cardinalities of duplicated training examples similar to the method used in \cref{subsec:hyperparam_anal}.

API vendors do not currently release information on how the training loss is computed, and thus, it can reduce the strength of the attacks, but as our experimental analysis has shown, the attacker does not actually need to fully reverse engineer the loss for it to be a useful signal for discrete prompt optimization.

\newparagraph{Mitigations that scan the training set.} Prior work in malicious fine-tuning discovered that vendors implement pre-fine-tuning moderation using classifiers that look for the presence of ``malicious'' data in the training file~\cite{halawi2024covert}. For example, jailbreaking content or particularly well-known prompt injection methods (\eg ``ignore previous instructions'') get flagged and blocked. However, this is not a complete defense because it can be evaded by encoding the training set to hide its purpose~\cite{halawi2024covert}.
\section{Related work}

Existing LLMs are vulnerable to various attacks. Prompt injections and jailbreaking are two types of attacks against LLMs that have attracted substantial attention from LLM vendors recently~\cite{gemmateam2024gemma2improvingopen}. Prompt injections, assuming partial control over the LLM input, aim to manipulate LLM to cause user-unintended behavior \eg tool misuse and data leakage~\cite{fu2023misusing,wunderwuzzi,greshake2023youve,promptinjection}. Jailbreaking, on the other hand, aims to get the LLM to respond to user requests that violate the safety policy specified by the LLM vendor \eg generating harassment content. Although these two types of attacks show distinct threat models (the user is benign in the former one while being adversarial in the latter one) and objectives, they both require manipulating the LLM into generating specific text desired by the attackers. To achieve this goal, there are manual and automated methods.

\newparagraph{Linguistic Prompt Injections.} Existing prompt injection attacks on real products are typically hand-crafted and exploit model-specific quirks~\cite{greshake2023youve,liu2023prompt,markdownattack,zhan2024injecagent, yi2023benchmarking}. For instance, ``Ignore previous instructions ..." effectively forces the LLM to follow subsequent instructions and disregard any ethical constraints placed beforehand~\cite{perez2022ignore}.
Other attacks such as \cite{pasquini2024neural} rely on creating a separation using long strings of delimiters. Such separation naturally allows the malicious instruction to stand apart from the prior context. Similarly, there are manually crafted jailbreaking attacks~\cite{li2023multistepjailbreakingprivacyattacks,anthropicManyshotJailbreaking}. Among them, Anil \etal explores using multi-shot in-context examples to bypass vendor-specified safety policies~\cite{anthropicManyshotJailbreaking}.  These attacks are ad-hoc and arguably easy to patch \eg blocking suspicious prompts such as the aforementioned ``Ignore previous instructions''~\cite{thevergeOpenAIsLatest}. Also, extending these attacks to new LLM products or new objectives usually involves redundant manual effort and thus is not suitable for systematic large-scale attackers. By contrast, our attack is principled and does not rely on prompt tinkering.

\newparagraph{Automated Prompt Injections.} Depending on the knowledge required about the LLM, automated methods are classified into whitebox ones, blackbox ones, and graybox ones. 

Whitebox methods require full access to model weights for the computation of gradients~\cite{zou2023universal,pasquini2024neural,sadasivan2024fastadversarialattackslanguage}. 
Greedy Coordinate Gradient is a pioneering whitebox algorithm originally designed for jailbreaking but can also work for prompt injections~\cite{zou2023universal}. It utilizes gradient information to guide the search for an adversarial input. NeuralExec uses the Greedy Coordinate Gradient algorithm to generate automated whitebox prompt injection  attacks~\cite{pasquini2024neural}. Due to the requirements of model weights, most of these whitebox attacks were evaluated on open-weight models.

In the blackbox setting, prior work has used other LLMs or natural language based heuristics to guide the search for jailbreak prompts\cite{chao2024jailbreakingblackboxlarge,liu2024autodangeneratingstealthyjailbreak}.

Finally, in the graybox setting, attacks do not need model weights but utilize other related information such as logprobs~\cite{hayase2024query,andriushchenko2024jailbreakingleadingsafetyalignedllms}. The logprobs based attacks rely on being able to compute the logprobs of a target token using sampling parameters such as "logit bias" which they can use to guide their search algorithm. However, LLM vendors which were vulnerable to such attacks have modified (or can easily otherwise restrict) their APIs from giving this information \cite{openaiapireference,geminiteam2024gemini15unlockingmultimodal}.

In contrast, our attack, as a graybox attack, proposes a novel attack channel --- the fine-tuning interface. This attack vector is hard to mitigate considering that fine-tuning loss is a critical component required by fine-tuning users. 

\newparagraph{Covert Malicious Fine-Tuning.} Other than adversarial prompts, LLMs can also be attacked by perturbing the model weights \cite{wan2023poisoning,zhao2023learning}. Models like OpenAI's GPT can be misaligned by finetuning on less than 100 malicious prompts \cite{zhan2023removing,qi2023fine}. However, users can finetune closed source models only via their finetuning APIs where the model provider can inspect the training data prior to finetuning. Recent work has proposed encoding the training data to covertly finetune on malicious data \cite{halawi2024covert}. Our work is orthogonal to this line of research as we don't rely on updating the model weights, rather we use the loss metrics reported during fine-tuning to gain more information about the base model.

\newparagraph{Reverse Engineering Closed-Weights LLMs.}
Model stealing is a well-studied problem where the adversary's goal is to extract model weights using only query access to the target model \cite{tramer2016stealing}. While model stealing is a more challenging task for larger models, the growing number of closed source LLMs has inspired attacks that extract more limited information. One class of attacks attempts to retrieve the exact dimension of hidden layers \cite{wei2020leaky, carlini2024stealingproductionlanguagemodel}. Others have tried to recover the total number of model parameters by correlating performance on benchmarks with results of open-source models \cite{gao2021sizes}. Similarly, attacks have tried to recover tokenizers of closed-source LLMs \cite{rando_anthropic_tokenizer_[Year]}. In this work, we partially reverse-engineer the workings of the closed-source Gemini fine-tuning API.
\section{Conclusion}

Our goal is to move towards safe and secure LLM systems. A pre-requisite for that is to thoroughly evaluate all the attack vectors that these emerging systems face. This helps focus defense efforts on threats that matter. Our work opens a new direction of investigation that analyzes the attack surface of remote fine-tuning interfaces. This is a popular and emerging feature in the LLM landscape and we provide the first adversarial analysis. We experimentally characterized the loss signal returned from the Google Gemini fine-tuning interface and showed how it can be used to create prompt injection attacks, through a simple discrete prompt optimization algorithm. Mitigating this attack vector is non-trivial because any restrictions on the training hyperparameters would reduce the utility of the fine-tuning interface. Arguably, offering a fine-tuning interface is economically very expensive (more so than serving LLMs for content generation) and thus, any loss in utility for developers and customers can be devastating to the economics of hosting such an interface. We hope our work begins a conversation around how powerful can these attacks get, and what mitigations strike a balance between utility and security.

\section*{Acknowledgements}
We thank our shepherd, the anonymous reviewers, Ilia Shumailov, Taylor Berg-Kirkpatrick, Ivan Evtimov, Cosmin Negruseri, Charles Staats, and the Geek Club. This work is supported in part by gifts from Amazon and Google and by NSF award 2312119.

\bibliographystyle{IEEEtran}
\bibliography{references}
\appendices

\section{}
\subsection{Provably correct method to recover permutations}
Given an input sequence of size $N$: $X = (x_1, x_2, ..., x_N)$. The fine-tuning API permutes this sequence before compute the training loss i.e. $FT(x_1,x_2,...,x_N) = (l_{\sigma_N(1)},l_{\sigma_N{2}},...,l_{\sigma_N(N)})$, where $l_i$ is the loss corresponding to $x_i$ and $\sigma_N: \{1,2,...,N\} \rightarrow \{1,2,...,N\}$ is the permutation function. Note than $\sigma_N$ is shuffles in a deterministic order depending on the value of $N$. Our goal is to recover $\sigma_N$ so that we can correctly ordered loss values.

\begin{prop} 
Given a permutation function $\sigma_{\sqrt{N}}$, an adversary can recover the permutation function $S_{N}$ by making 3 requests to the fine-tuning API.
\end{prop}
\begin{proof}
    Given an input sequence, $X = (x_1,...x_{\sqrt{N}})$, the adversary can get permuted losses $L' = (l_{S_{\sqrt{N}}(1)}, ..., l_{S_{\sqrt{N}}(\sqrt{N})})$ by making 1 fine-tuning request. Given access to the permutation function $S_{\sqrt{N}}$, it is trivial to recover the correct ordering $L = (l_1,...,l_{\sqrt{N}})$. Here, we assume that $l_i \neq l_j \forall i \neq j$.
    Now, let us construct a larger sequence of size $N$, $X_N = (x_1, x_1, ... \sqrt{N} \; \text{times}, x_2, x_2, ... \sqrt{N} \; \text{times}, ..., x_{\sqrt{N}}, ...)$. Now, by making the second fine-tuning request, the adversary can get $L'_N$. However, since the input had repeated values, the losses will also have repeated values. Particularly, it will have $\sqrt{N}$ instances each of $l_1$, $l_2$, ..., $l_{\sqrt{N}}$. 
    Now, let us construct another sequence of size $N$, $X'_N = (x_1,x_2,...,x_{\sqrt{N}}, x1, x2, ..., x_{\sqrt{N}}, ... \sqrt{N}\; \text{times})$. Finally, we make the third fine-tuning request to get $L''_N$. Now, it is easy to see that we can reconstruct $\sigma_N$ by using the loss values in $L'_N$ and $L''_N$. Concretely, 
    \[
    \sigma_N(i) = p \; | \; l'_p = l_{\lfloor \frac{i}{\sqrt{N}} \rfloor}, l''_p = l_{ i \% \sqrt{N}}
        \]
\end{proof}

It is trivial to get $S_2$ by making three calls to the fine-tuning API. Therefore, the above provable method has the complexity $\mathcal{O}(3 \log_2(\log_2(N)))$. In comparison, our method in \cref{subsec:recover_random} can get the permutation using only 1 fine-tuning request. 

In \cref{subsec:recover_random}, we described our method to recover the permutation of training losses reported by the fine-tuning API. For a training set of size $N$, our method only needs to make 1 fine-tuning request of size $N$. However, the approach only approximates the permutation since it relies on the assumption that progressively corrupted strings should result in increasing training loss values. In this section, we evaluate the accuracy of our method by comparing it against the alternate permutation recovering method that is query inefficient but provably correct under the weaker assumption that losses are unique.

To evaluate our approximate permutation recovery method, we compare our approximate method against this provably correct method as the ground truth. We compare the approximate permutation with the provably correct permutation using two comparison measures - 
\begin{newitemize}
    \item \textbf{Normalized Hamming distance}, that is, number of positions where the approximate permutation differs from the provably correct permutation, normalized by the length of the permutation
    \item \textbf{Kendall Correlation} which measures the fraction of pairwise orderings that are preserved between the approximate permutation and the provably correct permutation.
\end{newitemize}
Since our approximate algorithm is a randomized algorithm, we report the averages of the Normalized Hamming Distance and Kendall Correlations, averaged over $5$ approximate permutations. The results are as shown in \cref{tab:perm_comp_results}.

The Average Normalized Hamming distance numbers show that the approximate method misidentifies only a small subset of the permutation, and when it does, the relative orderings are still preserved to a high degree (as is shown by the high Kendall correlations)

\begin{table}[t]
  \centering
  \caption{The approximate permutations are close to the true permutation across a large range of training dataset sizes}
\begin{adjustbox}{width=\linewidth}
  \begin{tabular}{lcc }
    \toprule
    Training Dataset Size & Avg. Norm. Hamming Dist. (Std. Dev.) &  Avg. Kendall Corr. (Std. Dev.)\\

    \midrule
    100 & 0.036 (0.029) & 0.947 (0.035)\\
    200 & 0.062 (0.007) & 0.934 (0.019)\\
    300 & 0.061 (0.012) & 0.925 (0.014)\\
    400 & 0.064 (0.010) & 0.912 (0.017)\\
    500 & 0.060 (0.006) & 0.919 (0.004)\\
    600 & 0.067 (0.009) & 0.915 (0.013)\\
    700 & 0.074 (0.015) & 0.904 (0.031)\\
    800 & 0.066 (0.011) & 0.917 (0.026)\\
    900 & 0.0733 (0.022) & 0.898 (0.006)\\
    1000 & 0.0736 (0.010) & 0.905 (0.008)\\
    \bottomrule
  \end{tabular}
\end{adjustbox}
  \label{tab:perm_comp_results}
\end{table}

\subsection{Injection types for the PPL40 vs. Purle Llama}

Number of indirect prompt injections of each type for both PPL40 vs. Purle Llama is shown in \cref{tab:purplellama_dist_table}.

\begin{table}[h]
    \centering
     \caption{Number of indirect prompt injections of each type shows that our sampled PPL40 reflects the Purle Llama distribution}
   \begin{tabular}{c c c}
   \toprule
        Attack template & Purple Llama & PPL40 \\
        \midrule
        \textit{ignore prev. instructions} & 8 & 7 \\
         \textit{persuasion} & 6 & 6     \\
         \textit{different input language} & 7 & 4 \\
         \textit{system mode} & 7 & 7 \\
         \textit{hypothetical scenario} & 6 & 4 \\
         \textit{information overload} & 6 & 6 \\
         \textit{virtualization} & 6 & 5 \\
         \textit{token smuggling} & 6 & 0 \\
         \textit{mixed techniques} & 3 & 1\\
         \midrule
         \textbf{Total} & 55 & 40 \\
         \bottomrule
    \end{tabular}
    \label{tab:purplellama_dist_table}
\end{table}




\subsection{Fun-tuning ASR per prompt injection scenario}
\label{subsec:barplot_scenarios_methods_tables}
We show the attack success rate of our method classified by the kind of injection in \cref{fig:barplot_scenarios_methods_10_fig} and \cref{fig:barplot_scenarios_methods_15_fig}.

\begin{figure}[h]
    \centering
    \begin{tikzpicture}
    \begin{axis}[
        ybar,
        legend style={at={(0.5,0.95)},anchor=south,legend columns=-1},
        ylabel={ASR\%},
        ytick distance=25,
        y label style={at={(axis description cs:0.035,.5)}},
        xlabel={Scenario},
        x label style={at={(axis description cs:0.5,-.23)}},
        symbolic x coords={code,exercise,population,transaction,password,zubrowka,resume,employee},
        xtick=data,
        xticklabel style={rotate=45,anchor=north east,yshift=0.2cm,xshift=0.05cm},
        bar width=4pt,
        ymin=0,
        width=\linewidth,
        height=5cm,
        ymajorgrids=true,
        grid style=dashed,
        ]
    \addplot table [x=category, y expr=\thisrow{Baseline}*100,col sep=comma] {plots/gemini1_bycat.txt};
    \addplot table [x=category, y expr=\thisrow{Ablation}*100,col sep=comma] {plots/gemini1_bycat.txt};
    \addplot table [x=category, y expr=\thisrow{Fun-tuning}*100,col sep=comma] {plots/gemini1_bycat.txt};
    \legend{Baseline,Ablation,Fun-tuning}
    \end{axis}
    \end{tikzpicture}
    \caption{ASR of our attack methods against Gemini 1.0 Pro per scenario shows that the Fun-tuning achieves $>75\%$ ASR in each scenario except the `password` phishing scenario, suggesting the Gemini 1.0 Pro might be good at recognizing phishing attempts of some form}
    \label{fig:barplot_scenarios_methods_10_fig}
\end{figure}

\begin{figure}[h]
    \centering
    \begin{tikzpicture}
    \begin{axis}[
        ybar,
        legend style={at={(0.5,0.95)},anchor=south,legend columns=-1},
        ylabel={ASR\%},
        ytick distance=25,
        y label style={at={(axis description cs:0.035,.5)}},
        xlabel={Scenario},
        x label style={at={(axis description cs:0.5,-.23)}},
        symbolic x coords={code,exercise,population,transaction,password,zubrowka,resume,employee},
        xtick=data,
        xticklabel style={rotate=45,anchor=north east,yshift=0.2cm,xshift=0.05cm},
        bar width=4pt,
        ymin=0,
        width=\linewidth,
        height=5cm,
        ymajorgrids=true,
        grid style=dashed,
        ]
    \addplot table [x=category, y expr=\thisrow{Baseline}*100,col sep=comma] {plots/gemini15_bycat.txt};
    \addplot table [x=category, y expr=\thisrow{Ablation}*100,col sep=comma] {plots/gemini15_bycat.txt};
    \addplot table [x=category, y expr=\thisrow{Fun-tuning}*100,col sep=comma] {plots/gemini15_bycat.txt};
    \legend{Baseline,Ablation,Fun-tuning}
    \end{axis}
    \end{tikzpicture}
    \caption{ASR of our attack methods against Gemini 1.5 Flash per scenario shows that the Fun-tuning achieves $>50\%$ ASR in each scenario except the `password` phishing and code analysis, suggesting the Gemini 1.5 Pro might be good at recognizing phishing attempts of some form and became better at code analysis}
    \label{fig:barplot_scenarios_methods_15_fig}
\end{figure}

\subsection{Example of successful  prompt injection in the text summarization task}
\label{subsec:barplot_scenarios_methods_tables}
We show another example of our attack in \cref{fig:example-attack}

\begin{figure}[]
    \centering
    \begin{adjustbox}{width=\linewidth}
            \input{plots/teaserimg}
    \end{adjustbox}
    \caption{Example prompt injection with our method on Gemini 1.0 Pro. Perturbations that trigger the prompt injection are highlighted in red and the injection payload is highlighted in bolded.}
    \label{fig:example-attack}
\end{figure}

\newpage 


\clearpage
\section{Meta-Review}

The following meta-review was prepared by the program committee for the 2025
IEEE Symposium on Security and Privacy (S\&P) as part of the review process as
detailed in the call for papers.

\subsection{Summary}
This paper presents a novel approach to optimize prompt-based attacks against LLMs by abusing the fine-tuning functionality provided by the LLM owner. Specifically, the paper shows that, by carefully setting the fine-tuning hyperparameters, the fine-tuning loss can be used as a proxy for information about the log probabilities of specific outputs for a given input.

\subsection{Scientific Contributions}
\begin{itemize}
\item Identifies an Impactful Vulnerability
\item Provides a Valuable Step Forward in an Established Field
\end{itemize}

\subsection{Reasons for Acceptance}
\begin{enumerate}
\item The paper identifies a vulnerability that significantly improves the effectiveness of prompt injection attacks against real-world services.
\item The paper describes and overcomes several technical challenges in order to use the fine-tuning functionality for this purpose.
\item The paper provides a thorough evaluation including an ablation study to demonstrate that fine-tuning training loss can be a valuable signal for prompt injection optimization.
\end{enumerate}

\subsection{Noteworthy Concerns} 
\begin{enumerate} 
\item The attack has been demonstrated on a single service, so it is not yet known which other services might be vulnerable to this type of technique.
\end{enumerate}

\section{Response to the Meta-Review}
\begin{enumerate}
\item The authors welcome the raised noteworthy concern and encourage more studies to understand or to exclude the attack feasibility for other services.
\end{enumerate}

\end{document}